\documentclass[twocolumn,amsmath,amssymb,prd]{revtex4}

\def\al{\alpha}
\def\be{\beta}
\def\ga{\gamma}
\def\de{\delta}

\def\et{\eta}

\def\ka{\kappa}
\def\la{\lambda}

\def\si{\sigma}

\def\ph{\phi}
\def\vp{\varphi}
\def\ch{\chi}
\def\ps{\psi}
\def\om{\omega}

\def\La{\Lambda}

\def\Ph{\Phi}

\def\mn{{\mu\nu}}

\def\fr#1#2{{{#1} \over {#2}}}
\def\half{{\textstyle{1\over 2}}}

\def\frac#1#2{{\textstyle{{#1}\over {#2}}}}

\def\vev#1{\langle {#1}\rangle}

\def\lsim{\mathrel{\rlap{\lower4pt\hbox{\hskip1pt$\sim$}}
    \raise1pt\hbox{$<$}}}
\def\gsim{\mathrel{\rlap{\lower4pt\hbox{\hskip1pt$\sim$}}
    \raise1pt\hbox{$>$}}}
\def\sqr#1#2{{\vcenter{\vbox{\hrule height.#2pt
         \hbox{\vrule width.#2pt height#1pt \kern#1pt
         \vrule width.#2pt}
         \hrule height.#2pt}}}}

\def\pt#1{\phantom{#1}}

\def\sss{s^{\mu\nu}}
\def\ttt{t^{\ka\la\mu\nu}}

\def\sb{\overline{s}}
\def\tb{\overline{t}}
\def\ub{\overline{u}}

\def\stw{\tilde{s}}
\def\ttw{\tilde{t}}
\def\utw{\tilde{u}}

\def\vb#1#2{e_{#1}^{{\pt{#1}}#2}}
\def\ivb#1#2{e^{#1}_{{\pt{#1}}#2}}
\def\uvb#1#2{e^{#1#2}}

\def\barvb#1#2{\bar e_{#1}^{{\pt{#1}}#2}}

\newcommand{\beq}{\begin{equation}}
\newcommand{\eeq}{\end{equation}}
\newcommand{\bea}{\begin{eqnarray}}
\newcommand{\eea}{\end{eqnarray}}
\newcommand{\bit}{\begin{itemize}}
\newcommand{\eit}{\end{itemize}}
\newcommand{\rf}[1]{(\ref{#1})}

\begin{document}


\title{Gravity with explicit spacetime symmetry breaking \\
and the Standard-Model Extension}

\author{Robert Bluhm, Hannah Bossi, and Yuewei Wen}

\affiliation{
Physics Department, Colby College,
Waterville, ME 04901, USA
}

\begin{abstract}
The Standard-Model Extension (SME) is the general phenomenological framework used to investigate
Lorentz violation at the level of effective field theory.
It has been used to obtain stringent experimental bounds on 
Lorentz violation in a wide range of tests.
In the gravity sector of the SME, it is typically assumed that the spacetime symmetry 
breaking occurs spontaneously in order to avoid potential conflicts with the Bianchi identities.
A post-Newtonian limit as well as matter-gravity couplings in the SME have been developed 
and investigated based on this assumption.
In this paper, the possibility of using the SME to also describe gravity theories
with explicit spacetime symmetry breaking is investigated.
It is found that in a wide range of cases, 
particularly when matter-gravity couplings are included,
consistency with the Bianchi identities can be maintained,
and therefore the SME can be used to search for signals of the symmetry breaking.
Two examples with explicit breaking are considered.
The first is ghost-free massive gravity with an effective metric that couples to matter.
The second is Ho\v rava gravity coupled with matter in an infrared limit.
\end{abstract}


\maketitle

\section{Introduction}

General relativity (GR) and the Standard Model (SM) of particle physics
are well-tested theories that describe the fundamental forces of nature.
However, GR is not a quantum theory,
since it is not renormalizable, 
and it must be treated as an effective field theory at low energies.
This as well as open questions about the nature of dark matter and dark energy
have led to investigations of alternative gravity theories that modify GR,
where the ultimate goal is to find a consistent quantum theory of gravity.
In many scenarios,
small violations of local Lorentz and diffeomorphism invariance can occur,
which would provide important signatures of new physics
\cite{overviews}.

The phenomenological framework known as the Standard-Model Extension (SME)
has been developed and used to search for signals of spacetime symmetry breaking
in a wide range of experimental tests
\cite{sme,akgrav04,rbsme,aknr-tables}.
The Lorentz- and diffeomorphism-breaking operators that appear in the SME
involve couplings with fixed background fields, usually referred to as SME coefficients.
The results of experimental tests can be
interpreted as bounds on the SME coefficients.
Many different types of operators and SME coefficients have been classified
and probed. 
These include both power-counting renormalizable and nonrenormalizable operators
\cite{higherdimops}.
Gravity sectors in the SME can be defined using metric or vierbein descriptions in 
Riemann spacetime or more generally in Riemann-Cartan spacetime
\cite{akgrav04}.
Relationships between Lorentz violation and torsion
\cite{smetorsion},
nonmetricity
\cite{nonmetric}, 
and Riemann-Finsler geometry
\cite{fins} have been
explored using the SME.

In investigations involving gravity,
a post-Newtonian limit of the SME has been developed 
\cite{qbak06}
and matter-gravity interactions have been incorporated
\cite{akjt}.
These are used to examine a variety of experiments, including
lunar laser ranging tests
\cite{lrange}, 
atom interferometry
\cite{atomint},
short-range gravitational tests
\cite{srgrav},
analyses of baryon number asymmetry
\cite{GL06},
orbital motion analyses 
\cite{orb}, 
gyroscope precession
\cite{gyro},
pulsar timing
\cite{pulsar}, 
perihelion and solar-spin tests
\cite{qbak06,Iorio},
and analyses of gravitational \v Cerenkov radiation
\cite{smegceren}.
Lorentz-violating effects in gravitational radiation
have also been investigated using a linearized version
of the SME
\cite{smegwaves}.

In the gravity sector of the SME,
an important distinction is made between 
spontaneous and explicit spacetime symmetry breaking
\cite{akgrav04}.
With explicit breaking,
the SME coefficients are nondynamical background tensors,
and they appear directly in the Lagrangian as objects with 
preferred spacetime directions
\cite{rb15a,rb15b,rbas16}.
However, with spontaneous breaking,
all tensors are dynamical,
and the SME coefficients arise as vacuum expectation values
\cite{ks}.
With spontaneous breaking,
the usual Noether identities involving 
the Bianchi identities, the Euler-Lagrange equations for matter fields,
and covariant energy-momentum conservation all hold similarly 
to how they hold in GR.
In contrast, with explicit breaking,
potential conflicts can occur between the Bianchi
identities, the dynamical equations of motion, and covariant energy-momentum conservation.
In some cases,
this puts severe restrictions on a theory or
results in it being inconsistent.

It is for this reason that the spacetime symmetry breaking in the SME with gravity 
is usually assumed to be spontaneous,
since potential inconsistencies with the
Bianchi identities are then avoided
\cite{akgrav04}.
This is because with spontaneous breaking,
the excitations have a known form as massless Nambu-Goldstone (NG)
modes or as massive Higgs-like excitations,
and these together with the vacuum solutions combine in a way 
that maintains the unbroken spacetime symmetry.
The fact that the excitations have a known form also
plays an important role in developing the post-Newtonian limit of the SME. 
It allows for a systematic perturbative treatment that does
not depend on the particular structure of an underlying Lorentz-breaking theory,
and this in turn allows the post-Newtonian limit of the SME to be applied
in a wide range of experimental tests.

The goal of this paper is to take a closer look at the case of explicit spacetime
symmetry breaking and to show that in a wide range of cases the SME can still be applied. 
This requires looking at the types of interactions 
and the form of the extra excitations that can occur with explicit breaking
and showing that a useful post-Newtonian limit with consistent matter-gravity 
couplings can still be obtained.
As examples, the SME is used to investigate effects of
explicit spacetime symmetry breaking that might occur in
ghost-free massive gravity \cite{MGreviews}
and Ho\v rava gravity
\cite{HGreviews} when matter-gravity couplings are included.
Specifically, the first example looks at ghost-free massive gravity 
with matter couplings formed
using an effective potential consisting of a combination of the physical metric
or vierbein and a nondynamical background.
The second example considers possible matter couplings that might arise in
the infrared limit of Ho\v rava gravity.

The organization of this paper is as follows.
Section II provides background on local Lorentz and diffeomorphism breaking in gravity,
including a discussion of the differences between spontaneous and explicit breaking.
This is followed in Sec.\ III by a brief overview of the gravity sector of the SME.
Section IV looks at what happens in the gravity sector of the SME
when the symmetry breaking is explicit as opposed to spontaneous.
This is followed by an examination of two gravity models with 
explicit breaking in Section V.
A summary and conclusion are presented in Sec.\ VI.

\section{Gravity and Lorentz violation}

At the level of effective field theory,
local Lorentz symmetry and diffeomorphism invariance are broken
when matter and gravitational fields interact with a 
fixed background tensor that has preferred directions in spacetime.

For a theory of this type in Riemann spacetime,
the general form of the action can be written as
\bea
S = {\Huge \int} d^4x \sqrt{-g} \, \left[  {\cal L}_{\rm grav} (g_{\mu\nu}) +
{\cal L}_{\rm LI} (g_{\mu\nu},\vp_\si) \right.
\nonumber \\
+ \left. {\cal L}_{\rm LV} (g_{\mu\nu},\vp_\si, \bar k_{\mu\nu\cdots}) \right] .
\quad
\label{S1}
\eea
In this expression,
the components of the metric, $g_{\mu\nu}$, are defined with respect to a spacetime coordinate frame.
Symmetry-preserving pure-gravity terms are contained in ${\cal L}_{\rm grav} (g_{\mu\nu})$,
which is assumed to include an Einstein-Hilbert term.
Conventional tensor matter fields are denoted collectively as $\vp_\si$,
where $\si$ denotes all of the relevant indices.
The term ${\cal L}_{\rm LI} (g_{\mu\nu},\vp_\si)$ includes 
all Lorentz-invariant (LI) and diffeomorphism-invariant matter-gravity interactions.
The background field associated with the symmetry breaking is denoted 
with an unspecified number of indices as $\bar k_{\mu\nu\cdots}$.
The Lorentz-violating (LV) and diffeomorphism-breaking terms 
are all contained in ${\cal L}_{\rm LV} (g_{\mu\nu},\vp_\si, \bar k_{\mu\nu\cdots})$.

To generalize to Riemann-Cartan spacetime,
which allows fermions as well as torsion to be included,
a vierbein formalism is used.
The vierbein, $\vb \mu a$, has components defined with respect to both the 
spacetime frame and a local Lorentz frame,
and covariant derivatives involve a spin connection
\cite{akgrav04,rbak}.
For simplicity, only models in a zero-torsion limit with a spin connection 
defined entirely in terms of the vierbein are considered here.
In this case,
the generic form of the action can be written as
\bea
S = {\Huge \int} d^4x \, e \left[  {\cal L}_{\rm grav} (\vb \mu a) +
{\cal L}_{\rm LI} (\vb \mu a,\vp_\si,\ps) \right.
\nonumber \\
+ \left. {\cal L}_{\rm LV} (\vb \mu a,\vp_\si, \ps, \bar k_{ab\cdots}, \barvb \mu a) \right] .
\quad
\label{S2}
\eea
Here, $e$ is the determinant of the vierbein,
$\ps$ represents a generic fermion field,
and $\bar k_{ab\cdots}$ are the components of the background relative to
the local Lorentz frame.
A background vierbein, denoted as $\barvb \mu a$,
provides a link between the
spacetime and local frame components of the fixed background tensor.

Note that if a theory is defined initially by making tensor contractions of
the spacetime indices $\bar k_{\mu\nu\cdots}$ with only dynamical fields
in the spacetime frame,
then the background vierbein must be used to introduce the
local components $\bar k_{ab\cdots}$,
where these are related by
\beq
\bar k_{\mu\nu\cdots} = \barvb \mu a \barvb \mu a \cdots \bar k_{ab\cdots} .
\label{kmunuab}
\eeq
It is important to realize that if instead the components of $\bar k_{ab\cdots}$
are used directly to form contractions with dynamical fields in local frames,
then a different theory with different consequences and consistency conditions results.
See the Appendix for an illustration of this.

In both metric and vierbein descriptions,
all of the dynamical gravitational and matter fields transform appropriately 
under diffeomorphisms and local Lorentz transformations.
In contrast,
the components of tensor background fields remain fixed 
or transform anomalously under diffeomorphisms and local Lorentz transformations.
For example, the background vierbein $\barvb \mu a$ is fixed under both 
of these spacetime transformations.

It is important to emphasize that even though the background tensor 
can have preferred directions in spacetime,
a physically viable theory must still be observer independent.
This is a hallmark feature of the SME.
It requires that an effective field theory with spacetime symmetry 
breaking cannot depend on the choice of coordinates or local Lorentz basis.
The action and equations of motion must therefore be covariant 
under general coordinate transformations and under passive changes of local Lorentz bases.
Note that under these observer transformations, 
the components of the background tensor and vierbein transform, 
along with the gravitational and matter fields, 
in the conventional way.
An observer-independent Lagrangian can then be formed
as a scalar under the observer spacetime transformations.

\subsection{Spontaneous versus explicit breaking}

To understand the properties of the background tensor, 
it is necessary to make distinctions between spontaneous and 
explicit spacetime symmetry breaking
\cite{rb15a,rbas16}.

In the case of spontaneous breaking,
it is assumed that
the background tensor originates as the vacuum expectation 
value of a fully dynamical tensor.
The dynamical tensor can be denoted (depending on the frame) as
either $k_{\mu\nu\cdots}$ or $k_{ab\cdots}$.
These components are linked by the physical vierbein $\vb \mu a$.
The fixed background is then given as the vacuum expectation value,
e.g.,
\beq
\bar k_{\mu\nu\cdots} = \vev{k_{\mu\nu\cdots}} ,
\label{vevk}
\eeq
in the spacetime frame.
The components $\bar k_{ab\cdots}$ give the corresponding vacuum solution 
in the local Lorentz frame when a vierbein treatment is used,
and the background vierbein $\barvb \mu a$ is the vacuum expectation value
$\vev{\vb \mu a}$ of the physical vierbein $\vb \mu a$.

The full dynamical tensor can then be written as a sum of the
background plus excitations about the background,
where tildes are used to denote the excitations.
In the spacetime frame, 
this gives the expression
\beq
k_{\mu\nu\cdots} = \bar k_{\mu\nu\cdots} + \tilde k_{\mu\nu\cdots} ,
\label{kexpan}
\eeq
while in the local Lorentz frames, 
$k_{ab\cdots} = \bar k_{ab\cdots} + \tilde k_{ab\cdots}$.
Similarly, with spontaneous breaking, the physical vierbein is the sum of
the vacuum value plus excitations about the vacuum
\cite{rbak}.

With spontaneous symmetry breaking (SSB),
a complete dynamical description requires additional
kinetic terms in the action that describe the excitations 
as well as potential terms 
that induce the symmetry breaking.
Such terms can be denoted generically as 
${\cal L}_{\rm SSB}^{\prime} (g_{\mu\nu}, \vp_\si, \bar k_{\mu\nu\cdots}, \tilde k_{\mu\nu\cdots})$
in a metric description and can be added to the action in Eq.\ \rf{S1}.
Alternatively, in a vierbein description,
the additional terms would have the form
${\cal L}_{\rm SSB}^{\prime} (\vb \mu a, \vp_\si, \ps, \bar k_{ab\cdots}, \tilde k_{ab\cdots},\barvb \mu a)$
and would be added to Eq.\ \rf{S2}.

With SSB, 
the excitations $\tilde k_{\mu\nu\cdots}$ or $\tilde k_{ab\cdots}$ take the form
of NG modes and massive Higgs-like modes
\cite{rbak}.
The NG modes are generated by the broken symmetries.
For example, when diffeomorphisms generated by a vector $\xi^\mu$ are spontaneously broken,
the four infinitesimal NG excitations take the form of a Lie derivative 
acting on the vacuum solution $\bar k_{\mu\nu\cdots}$,
\beq
\tilde k_{\mu\nu\cdots} \simeq {\cal L}_\xi \bar k_{\mu\nu\cdots} .
\label{NGk}
\eeq
In this case, the Lie derivative can be expanded, 
and the vectors $\xi^\mu$ can be promoted to fields $\pi^\mu$ representing the NG modes.
The full dynamical tensor then has the form
\bea
k_{\mu\nu\cdots} \simeq 
&&
\bar k_{\mu\nu\cdots} 
+ (D_\mu \pi^\al) \bar k_{\al\nu\cdots}
+ (D_\nu \pi^\al) \bar k_{\mu\al\cdots}
+ \cdots
\nonumber \\
&&+ \pi^\al D_\al \bar k_{\mu\nu\cdots}
+ (\de k_{\mu\nu\cdots})_{\rm massive} ,
\label{diffNGmodes}
\eea
where $D_\mu$ represents covariant derivatives in the curved spacetime,
and the excitations
labeled as $(\de k_{\mu\nu\cdots})_{\rm massive}$
represent the massive Higgs-like modes that
generally occur with spontaneous symmetry breaking.

With SSB,
equations of motion for the dynamical tensor hold
when all the excitations are included.
These can be obtained as field variations on the action,
which have the form
\beq
\fr {\de {S}} {\de k_{\mu\nu\cdots}} = 0 .
\eeq
The background fields $\bar k_{\mu\nu\cdots}$ by themselves are the vacuum solutions,
which obey
\beq
\left( \fr {\de {S}} {\de \bar k_{\mu\nu\cdots}} \right)_{\rm vacuum} = 0 .
\eeq
Hence, with SSB the backgrounds $\bar k_{\mu\nu\cdots}$ are dynamical fields
in the sense that they satisfy the vacuum equations of motion.

In contrast, with explicit breaking the background tensor is nondynamical.
The components $\bar k_{\mu\nu\cdots}$ or $\bar k_{ab\cdots}$ are not vacuum values,
and there are no field variations with respect to them that
yield equations of motion.
Instead, mathematical variations with respect to $\bar k_{\mu\nu\cdots}$
result in expressions that need not vanish,
e.g.,
\beq
\fr {\de {S}} {\de \bar k_{\mu\nu\cdots}} \ne 0 .
\label{dSne0}
\eeq
With explicit breaking,
the background tensor is simply a prescribed nondynamical 
object that appears directly in the Lagrangian.

However, there are additional degrees of freedom that can appear
in a theory with explicit symmetry breaking.
This is because when diffeomorphisms and local Lorentz symmetry are broken explicitly,
there are fewer gauge freedoms.
These gauge freedoms would normally be used to eliminate components of the metric or vierbein,
but with explicit breaking these components instead remain as possible extra modes.
It is important to note as well that when a gauge symmetry is broken
the constraint structure of a theory is usually altered \cite{rbngrpav}.
This can further modify the nature and behavior of the physical degrees of freedom,
or it can cause a theory to be unphysical due to the appearance of ghost modes.
For this reason, it is important to work with theories where mechanisms exist
to eliminate potential ghosts.

\subsection{St\"uckelberg approach and explicit breaking}

It is common in theories with explicit symmetry breaking to use a St\"uckelberg approach
to describe the behavior of the extra degrees of freedom that arise
\cite{ags03}.
In this approach, scalars are added as dynamical fields,
which restore the broken spacetime symmetry.
For example, with explicit diffeomorphism breaking in Riemann spacetime, 
four scalars, $\Ph^A$, with $A = 0,1,2,3$, are used to replace the background as follows:
\beq
\bar k_{\mu\nu\cdots}(x) \rightarrow  D_\mu \Ph^A D_\nu \Ph^B \cdots \bar k_{AB\cdots} (\Ph) .
\label{Stuck1}
\eeq
While this adds four extra degrees of freedom to the theory,
four local gauge freedoms (the restored diffeomorphisms) are created as well,
and therefore the net number of degrees of freedom remains unchanged.

The original theory with explicit breaking can be obtained from the
St\"uckelberg model by fixing the diffeomorphism
invariance so the four scalars match the spacetime coordinates:
\beq
\Ph^A = \de^A_\mu x^\mu .
\label{Phxgauge}
\eeq
Inserting this into \rf{Stuck1} gives back the original fixed background.
Notice, however, that if infinitesimal excitations about the coordinates are included 
in the St\"uckelberg scalars,
denoted as fields $\pi^\mu$, where
\beq
\Ph^A = \de^A_\mu (x^\mu + \pi^\mu) ,
\label{Phexcite}
\eeq
then an expansion in Eq.\ \rf{Stuck1} gives:
\bea
D_\mu \Ph^A D_\nu \Ph^B \cdots \bar k_{AB\cdots} (\Ph) 
\simeq 
\bar k_{\mu\nu\cdots} 
+ (D_\mu \pi^\al) \bar k_{\al\nu\cdots}
\nonumber \\
+ (D_\nu \pi^\al) \bar k_{\mu\al\cdots}
+ \cdots
+ \pi^\al D_\al \bar k_{\mu\nu\cdots} .
\quad\quad
\label{Stuckmodes}
\eea
Comparing this with Eq.\ \rf{diffNGmodes} shows that the infinitesimal excitations
in the St\"uckelberg approach reproduce the NG excitations that would occur in a
similar theory with spontaneous breaking.  

While the infinitesimal NG modes are found to be the same,
there are still some important differences that remain between an explicit-breaking
theory with St\"uckelberg fields and a theory with spontaneous breaking.
For example,
there are still no dynamical field equations for $k_{\mu\nu\cdots}$ in the
St\"uckelberg approach,
and there are no massive Higgs-like excitations 
$(\de k_{\mu\nu\cdots})_{\rm massive} $
in \rf{Stuckmodes} as there are in \rf{diffNGmodes}.
There are also additional terms that would appear in the action of a
theory with spontaneous breaking, such as ${\cal L}_{\rm SSB}^{\prime}$,
which are absent in an explicit-breaking model.
What the St\"uckelberg approach does is it introduces the minimal number of
excitations that are needed to restore the broken symmetry,
which is four in the case of broken diffeomorphisms,
and it does so by creating the same NG modes 
that would appear in a theory with spontaneous breaking.

\section{Gravity and the SME}

The SME is constructed as the general observer-independent effective field theory
formed from matter and gravitational fields interacting with Lorentz-violating tensors.
The theory contains the SM and GR,
including possible Lorentz-preserving extensions,
as well as a multitude of additional interaction terms that lead
to breaking of spacetime symmetry.

Typically experiments test for signatures of Lorentz breaking and express their
results as limits or bounds on the SME coefficients.
In most investigations in Minkowski spacetime,
the SME coefficients are treated as constant to first approximation
(see \cite{cl16} for an analysis including time dependence).
As a result,
global translation invariance still holds,
while global Lorentz symmetry is broken.
In this context, it is not crucial whether the SME coefficients are viewed
as vacuum expectation values or purely as phenomenological coefficients.

However, when gravity is included and GR becomes a limiting
subsector of the SME,
there are geometrical constraints, such as the Bianchi identities, 
which become important.
Moreover, with gravity, Lorentz symmetry becomes a local symmetry,
and diffeomorphism invariance appears as an additional local symmetry.  
Field theories with local symmetries have associated Noether identities
that link the Euler-Lagrange equations obeyed by the 
dynamical fields in the theory.
In GR, for example,
the divergence of the Einstein equations, $G^{\mu\nu} = 8 \pi G T^{\mu\nu}$
is linked off shell via Noether identities to the Euler-Lagrange equations for the
dynamical matter fields.
When the matter fields are on shell,
and the contracted Bianchi identity, $D_\mu G^{\mu\nu} = 0$,
is used, the result is that $D_\mu T^{\mu\nu} = 0$ holds automatically
as a result of the identities.
Essentially, the four diffeomorphism invariances in GR
cause the four equations $D_\mu T^{\mu\nu} = 0$ to be redundant with
the Euler-Lagrange equations for the dynamical matter fields.

These relations between the Bianchi identities, 
the dynamical Euler-Lagrange equations
and covariant energy-momentum conservation continue to hold even 
when spontaneous spacetime symmetry breaking occurs,
since the fields are all dynamical.
However, if a nondynamical tensor field is introduced,
which explicitly breaks spacetime symmetries,
it no longer has to obey Euler-Lagrange equations,
and as a result potential inconsistencies with the Bianchi identities can arise.
It is for this reason that the SME coefficients are typically assumed to arise as
a result of spontaneous local Lorentz and diffeomorphism breaking.

\subsection{Minimal SME with gravity}

The minimal SME in Riemann spacetime 
restricts the Lorentz-breaking operators to dimension four or less,
and it traditionally assumes the spacetime symmetry breaking is spontaneous.
The resulting action can be divided into sectors,
\beq
S_{\rm SME} \simeq \int d^4x \sqrt{-g} [ \fr 1 {16 \pi G} R + {\cal L}_{\rm LV} + {\cal L}_{\rm LI} + {\cal L}_{\rm SSB}^{\prime}]. 
\label{Sparts}
\eeq
The terms in ${\cal L}_{\rm LV}$ contain the diffeomorphism and Lorentz violating 
interactions of the SME coefficients with gravitational and matter fields,
while the ordinary symmetry-preserving matter terms, 
including their couplings to gravity, 
are in ${\cal L}_{\rm LI}$.
The terms in ${\cal L}_{\rm SSB}^{\prime}$ contain the dynamical terms for the excitations of the
SME tensors that occur in a process of spontaneous symmetry breaking.

At leading order in the SME coefficients,
${\cal L}_{\rm LV}$ can be divided into terms with pure-gravity 
and matter-gravity couplings,
\beq
{\cal L}_{\rm LV} \simeq {\cal L}_{\rm LV}^{\rm (grav)} + {\cal L}_{\rm LV}^{\rm (matter-grav)}
\label{LLV}
\eeq
The pure-gravity couplings at this level of approximation involve three 
interaction terms given as
\beq
{\cal L}_{\rm LV}^{\rm (grav)}  = \fr 1 {16 \pi G} \left( -u R 
+\sss R^T_\mn + \ttt C_{\ka\la\mu\nu} \right) ,
\label{llvmsme}
\eeq
where
$R^T_\mn$ is the trace-free Ricci tensor 
and $C_{\ka\la\mu\nu}$ is the Weyl conformal tensor.  
The fields $s^\mn$ and $t^{\ka\la\mu\nu}$
have symmetries that match those of
the trace-free Ricci tensor and the Riemann curvature tensor,
respectively.  

The dynamical fields $u$, $s^\mn$ and $t^{\ka\la\mu\nu}$
give rise to SME coefficients
$\bar u$, $\bar s^\mn$ and $\bar t^{\ka\la\mu\nu}$
as vacuum values in a process of spontaneous local 
Lorentz and diffeomorphism breaking.
This permits a separation of the dynamical fields into SME coefficients
and small fluctuations denoted using tildes,
\bea
u &=& \ub + \utw,
\nonumber\\
\sss &=& \sb^\mn + \stw^\mn,
\nonumber\\
\ttt &=& \tb^{\ka\la\mu\nu} + \ttw^{\ka\la\mu\nu}.
\label{texp}
\eea
Since the SME coefficients originate from spontaneous symmetry breaking,
the excitations $\utw$, $\stw^\mn$, and $\ttw^{\ka\la\mu\nu}$ 
consist of NG modes and massive Higgs-like modes,
where the terms in the action describing these excitations are contained in 
${\cal L}_{\rm LV}$ and ${\cal L}_{\rm SSB}^{\prime}$.
While there may not be known expressions for  
${\cal L}_{\rm LV}$ and ${\cal L}_{\rm SSB}^{\prime}$,
the consistency of the theory is assured, since $D_\mu T^{\mu\nu} = 0$ holds automatically
as the result of the Noether and Bianchi identities
when the excitations $\utw$, $\stw^\mn$, and $\ttw^{\ka\la\mu\nu}$ are on shell.

In applications where gravity is weak,
the metric can be expanded perturbatively about a Minkowski background,
$g_{\mu\nu} \simeq \et_{\mu\nu} + h_{\mu\nu}$,
and the effects of gravity in a post-Newtonian limit can be investigated.
The post-Newtonian limit of the SME is described in Ref.\ \cite{qbak06},
where a systematic procedure based on a general set of 
assumptions is used to find an expansion that
decouples the fluctuations, $\tilde u$, $\tilde s^\mn$, and $\tilde t^{\ka\la\mu\nu}$,
from the vacuum values and metric excitations.  
Central to this procedure is the fact that with spontaneous symmetry breaking,
diffeomorphism invariance holds and consistency of the dynamics
with covariant energy-momentum conservation is maintained.
The result is a post-Newtonian description
involving only the metric and the SME coefficients.  
Interestingly, in this context, sensitivity to $\bar u$ and $\bar t^{\ka\la\mu\nu}$ in
these expansions does not appear
\cite{qbak06,akjt,yb15};
however, cosmological inflationary models may have
effects depending on $\bar t^{\ka\la\mu\nu}$
\cite{yb17}.
It is for this reason that the coefficients $\bar u$ and $\bar t^{\ka\la\mu\nu}$
are largely ignored in the remainder of this paper,
including in Sec.\ V.
The bounds obtained for the pure-gravity sector of the minimal SME
only involve the $\bar s^\mn$ coefficients.

Matter-gravity couplings in the minimal SME have been analyzed as well
and are described in Ref.\ \cite{akjt}.
In this case,
a systematic perturbative method is developed,
and it is used to investigate Lorentz-violating
effects involving matter particles or light in a weak gravitational field.
With matter included,
the minimal SME terms in ${\cal L}_{\rm LV}^{\rm (matter-grav)}$ 
include a number of coefficients that couple with gravity.
For example, 
a fermion has couplings with coefficients $a_\mu$, $b_\mu$, $c_{\mu\nu}$,
$d_{\mu\nu}$, $e_\mu$, $f_\mu$, $g_{\la\mu\nu}$, and $H_{\mu\nu}$,
while a photon has couplings with coefficient $(k_F)^{\ka\la\mu\nu}$.
However, for the purposes of this paper, 
and in particular with regard to the examples considered in Sec.\ V,
it suffices to consider a subset of the SME matter-gravity couplings.
These are chosen to consist of a single fermion field $\ps$ 
coupled to a symmetric coefficient $c_{\mu\nu}$
and a photon field $A_\mu$ coupled to a coefficient $(k_F)^{\ka\la\mu\nu}$
having the same symmetries as the Riemann curvature tensor.

The relevant terms in this illustrative model are then given as
\bea
{\cal L}_{\rm LV}^{\rm (matter-grav)} 
&=& - \ivb \mu a \bar \ps c_{\al\be} \uvb \be a \ivb \al b \ga^b D_\mu \ps 
\nonumber \\
&& - \fr 1 4 (k_F)^{\ka\la\mu\nu} F_{\ka\la} F_{\mu\nu} ,
\label{Lps}
\eea
where a vierbein description is used due to the presence of the fermion.
Note that the covariant derivative reduces to partial derivatives in $F_{\mu\nu}$,
while it is given as
\beq
D_\mu \ps = \partial_\mu \ps + \fr 1 4 i \om_\mu^{\pt{\mu}ab} \si_{ab} \ps 
\label{Dps}
\eeq
when acting on a fermion field.
The dynamical tensors in the matter sector can be separated into background values
assumed to originate from spontaneous breaking plus fluctuations:
\bea
c_{\mu\nu} &=& \bar c_{\mu\nu} + \tilde c_{\mu\nu} ,
\nonumber \\
(k_F)^{\ka\la\mu\nu} &=& (\bar k_F)^{\ka\la\mu\nu} + (\tilde k_F)^{\ka\la\mu\nu} .
\label{cFflucs}
\eea
It is the backgrounds $\bar c_{\mu\nu}$ and $(\bar k_F)^{\ka\la\mu\nu}$ that
are probed at leading order in experimental tests.

\subsection{Field redefinitions}

As described in
Refs.\ \cite{akgrav04,akjt,yb15,dcpm02},
not all of the SME coefficients in a given experimental setup
are independent or physical.
In many cases, coordinate changes, component mixing in spinor space,
or field redefinitions can be used to move some of the sensitivity 
to Lorentz breaking from one particle sector to another,
or to remove a particular set of coefficients completely.
In particular,
in the presence of gravity, there are ten components of the SME 
coefficients that are not physical.
This can be seen in certain circumstances
as a direct result of having four coordinates and six local Lorentz
bases to choose.
Alternatively, the coordinates and bases can be left unchanged
while field redefinitions on the ten components in the metric
can be made that eliminate components of the SME coefficients.

To illustrate this,
consider a fermion of mass $m$ and a photon field in gravity,
where the Lorentz-violating tensors in the minimal SME are limited to 
$u$, $s^{\mu\nu}$, $c_{\mu\nu}$, and $(k_F)^{\ka\la\mu\nu}$.
The action including the usual Lorentz-invariant terms can then be written as
\bea
S_{\rm SME} &\simeq& 
\int d^4x \, e [ \fr 1 {16 \pi G} [ (1-u)R + s^{\mu\nu} R_{\mu\nu} ]
\nonumber \\
&& + i  \ivb \mu a \bar \ps ( \ga^a - c_{\al\be} \uvb \be a  \ivb \al b \ga^b) D_\mu \ps 
- m \bar \ps \ps
\nonumber \\
&& - \fr 1 4 F_{\ka\la} (g^{\ka\mu} g^{\la\nu} + (k_F)^{\ka\la\mu\nu}) F_{\mu\nu} ] ,
\label{sckF}
\eea
where in a vierbein treatment $g^{\mu\nu} = \ivb \mu a \ivb \nu b \et^{ab}$,
and the curvature tensor and covariant derivatives are derived using $\vb \mu a$.

In a perturbative approach that keeps terms to linear order in the fields 
$u$, $s^{\mu\nu}$, $c_{\mu\nu}$, and $(k_F)^{\ka\la\mu\nu}$,
it has been shown that redefinitions of the metric and vierbein can be used to eliminate dependence
on either $s^{\mu\nu}$ or $c_{\mu\nu}$ in $S_{\rm SME}$, 
or alternatively the symmetric combinations $(k_F)^{\al\mu\pt{\al}\nu}_{\pt{\al\mu}\al}$
can be eliminated
The new redefined metric is denoted as $\tilde g^{\mu\nu}$,
and it is related to the original metric by
\beq
g^{\mu\nu} \simeq (1+u) \tilde g^{\mu\nu} + s^{\mu\nu} ,
\label{tildeg}
\eeq
with $s^{\mu\nu}$ symmetric and traceless.
The new redefined vierbein is denoted as
$\tilde e^\mu_{\pt{\mu}a}$, 
and it is given by
\beq
e^\mu_{\pt{\mu}a} \simeq (1+ \half u) \tilde e^\mu_{\pt{\mu}a} 
+ \half \tilde e_{\si a}  s^{\mu\si}  .
\label{tildee}
\eeq
With these definitions, 
the following three integral relations have been shown to hold 
to first order in the SME coefficients
\cite{yb15}.

For the Einstein-Hilbert term,
\bea
\int d^4x \, \sqrt{-g} \fr 1 {16 \pi G} R 
\quad\quad\quad\quad\quad\quad\quad\quad\quad\quad\quad
\nonumber \\
\simeq \int d^4x \, \sqrt{-\tilde g} \fr 1 {16 \pi G} [(1-u) \tilde R +s^{\mu\nu} \tilde R_{\mu\nu}] .
\label{EHtilde}
\eea
Here, the curvature on the right-hand side,
which is defined in terms of the redefined metric $\tilde g^{\mu\nu}$,
is denoted with a tilde.
Note that a total derivative term in the integral on the right has been dropped.

For the Maxwell term,
\bea
\int d^4x \, \sqrt{-g}  \left( - \fr 1 4 F_{\ka\la} g^{\ka\mu} g^{\la\nu} F_{\mu\nu} \right) 
\quad\quad\quad\quad\quad\quad\quad\quad
\nonumber \\
\simeq \int d^4x \, \sqrt{-\tilde g} \left( - \fr 1 4 F_{\ka\la} (\tilde g^{\ka\mu} \tilde g^{\la\nu} + (k_F)^{\ka\la\mu\nu}) F_{\mu\nu} \right) \,
\label{Maxwelltilde}
\eea
where the SME coefficients are given as
\beq
(k_F)^{\ka\la\mu\nu} \simeq
s^{\ka [ \mu} \tilde g^{\nu ] \la} - s^{\la [ \mu} \tilde g^{\nu ] \ka} ,
\label{kFdef}
\eeq
which have as their symmetric trace
\beq
(k_F)^{\al\mu\pt{\al}\nu}_{\pt{\al\mu}\al} \simeq s^{\mu\nu} .
\label{symkF}
\eeq

Finally, for the fermion term
\bea
\int d^4x \, e [ i  e^\mu_{\pt{\mu}a} \bar \ps \ga^a D_\mu \ps - m \bar \ps \ps ]
\quad\quad\quad\quad\quad\quad
\nonumber \\
\simeq
\int d^4x \, \tilde e [ i  \tilde e^\mu_{\pt{\mu}a} \bar \ch \ga^a \tilde D_\mu \ch - m \bar \ch \ch
\quad\quad\quad
\nonumber \\
- i \tilde e^\mu_{\pt{\mu}a} \bar \ch \, c_{\al\be} \, \tilde e^{\be a}  \tilde e^\al_{\pt{\mu}b}  \ga^b \tilde D_\mu \ch ] 
\label{sckF2}
\eea
where the SME coefficients in this case are
\beq
c_{\mu\nu} = - \half (u \tilde g_{\mu\nu} + \tilde g_{\al\mu} \tilde g_{\be\nu} s^{\al\be} ) .
\label{calbedef}
\eeq
Notice that the covariant derivative defined using $\tilde e^\mu_{\pt{\mu}a}$ in \rf{sckF2}
is labeled with a tilde, 
and a rescaling of the fermion field $\ps$, relabeled as $\ch$, 
has been performed to keep the action in a standard Dirac form \cite{yb15}.

As Eqs.\ \rf{EHtilde}, \rf{Maxwelltilde}, and \rf{sckF2} reveal,
field redefinitions of the metric allow one set of the SME coefficients 
$s^{\mu\nu}$, $(k_F)^{\al\mu\pt{\al}\nu}_{\pt{\al\mu}\al}$, or $c_{\mu\nu}$ to be 
eliminated while altering the others.
For example, redefinitions can be made that eliminate $c_{\al\be}$ while
redefining $s^{\mu\nu}$ and $(k_F)^{\al\mu\pt{\al}\nu}_{\pt{\al\mu}\al}$, 
or that eliminate $(k_F)^{\al\mu\pt{\al}\nu}_{\pt{\al\mu}\al}$ 
while redefining $s^{\mu\nu}$ and $c_{\mu\nu}$.
The result is that one set of the components 
$s^{\mu\nu}$, $c_{\mu\nu}$, and $(k_F)^{\al\mu\pt{\al}\nu}_{\pt{\al\mu}\al}$
is unphysical.

Note, however, that if $u=0$  and $s^{\mu\nu}$, $c_{\mu\nu}$, and $(k_F)^{\ka\la\mu\nu}$ are
all related to a common traceless symmetric tensor $k^{\mu\nu}$ as
\bea
s^{\mu\nu} &\simeq& k^{\mu\nu}
\nonumber \\
(k_F)^{\ka\la\mu\nu} &\simeq& 
k^{\ka [ \mu} g^{\nu ] \la} - k^{\la [ \mu} g^{\nu ] \ka} 
\nonumber \\
c_{\mu\nu} &\simeq& - \half g_{\mu\al} g_{\nu \be} k^{\al\be} 
\label{metrickshift}
\eea
then all three sets of components $s^{\mu\nu}$, $c_{\mu\nu}$, and $(k_F)^{\al\mu\pt{\al}\nu}_{\pt{\al\mu}\al}$
can be eliminated by redefining the metric and vierbein.
Thus, in this special case where $u=0$ and the remaining
SME coefficients all originate from a common set of coefficients, 
which couple universally to gravity and all particle species in the same way as the metric,
there is no physical spacetime symmetry breaking.
In order to have physical and potentially measurable spacetime symmetry breaking,
at least one set of the coefficients 
$s^{\mu\nu}$, $c_{\mu\nu}$, and $(k_F)^{\al\mu\pt{\al}\nu}_{\pt{\al\mu}\al}$
must be independent of the others.

As a consequence of these field redefinitions,
experiments aimed at testing spacetime symmetry breaking in
matter-gravity interactions must have sensitivity to more than just one particle sector.
In the action given in \rf{sckF},
this requires that sensitivity to more than
one set of the coefficients $s^{\mu\nu}$, $c_{\mu\nu}$, and $(k_F)^{\al\mu\pt{\al}\nu}_{\pt{\al\mu}\al}$
must be attained.
Experiments that do achieve sensitivity to two or more of these sectors
can choose as a convention to set one set of coefficients to zero and place
bounds on a second independent set,
or they can define parameters consisting of combinations of SME coefficients 
from different sectors and place bounds on them.
For example,
atom interferometry tests that have sensitivity to both the gravity and
electromagnetic sectors of the SME,
place bounds on combinations of $s^{\mu\nu}$ and the coefficients 
contributing to $(k_F)^{\al\mu\pt{\al}\nu}_{\pt{\al\mu}\al}$.
In the standard Sun-centered celestial equatorial frame \cite{aknr-tables},
where indices are labeled using letters $JK\cdots$,
these combined parameters are denoted as $\si^{JK}$,
and measured bounds of order $10^{-9}$ have been placed on them
\cite{atomint}.

\section{SME with explicit breaking}

The question of whether the SME can be used to investigate gravity theories
with explicit local Lorentz and diffeomorphism breaking can be 
addressed generically using $\bar k_{\mu\nu\cdots}$
to represent an SME coefficient.
In this case, the background is assumed to be a fixed nondynamical tensor that
does not arise as a vacuum value from spontaneous spacetime symmetry breaking.
Instead, $\bar k_{\mu\nu\cdots}$ is to be interpreted as an SME coefficient 
that explicitly breaks spacetime symmetry.

A general form of the explicit-breaking action in a metric formalism can then be written as
\bea
S &=& \int d^4 x \sqrt{-g}  [ \fr 1 {16 \pi G} R + \,  {\cal L}_{\rm LI}(g_{\mu\nu}, \vp_\si)
\nonumber \\
&&+{\cal L}_{\rm R,\bar k}(R^\ka_{\pt{\ka}\la\mu\nu}, g_{\mu\nu}, \bar k_{\mu\nu\cdots})
\nonumber \\
&&+ {\cal L}_{\rm g,\bar k}(g_{\mu\nu}, \bar k_{\mu\nu\cdots}))
\nonumber \\
&&+ {\cal L}_{\vp,\bar k}(g_{\mu\nu}, \vp_\si, \bar k_{\mu\nu\cdots}) ] .
\label{explL}
\eea
This divides the action into distinct terms,
consisting of the Einstein-Hilbert term, 
a Lorentz-invariant matter term ${\cal L}_{\rm LI}$,
a term ${\cal L}_{\rm R,\bar k}$ containing contractions of the curvature tensor 
with the metric and the background,
a potential term ${\cal L}_{\rm g,\bar k}$
where the metric interacts with the background,
and a Lorentz-violating matter term ${\cal L}_{\vp,\bar k}$
involving interactions of the background with the metric and dynamical matter fields $\vp_\si$.
Note that some of these terms contain dimensional couplings.
The term ${\cal L}_{\rm g,\bar k}$ can in principle include covariant derivatives
acting on the background.
Note as well that each of these terms is assumed to be covariant under general coordinate transformations,
and thus each term in the Lagrangian is an observer scalar.
There could, of course, also be terms in which $\bar k_{\mu\nu\cdots}$,
$R^\ka_{\pt{\ka}\la\mu\nu}$, $\vp_\si$, and the metric all interact together,
but these are considered as sub-leading-order 
interaction terms compared to the ones given here.

Each of the Lagrangian terms in \rf{explL}
has a corresponding contribution to the energy-momentum tensor
obtained by varying the action with respect to the metric.
These can be written as $T_{\rm LI}^{\mu\nu}$, $T_{\rm R,\bar k}^{\mu\nu}$, $T_{\rm g,\bar k}^{\mu\nu}$, 
and $T_{\vp,\bar k}^{\mu\nu}$.
Assuming the Lorentz-invariant matter sector has a covariantly conserved 
energy-momentum tensor by itself,
consistency of the theory with the Bianchi identities requires that
\beq
D_\mu (T_{\rm R,\bar k}^{\mu\nu} + T_{\rm g,\bar k}^{\mu\nu} + T_{\vp,\bar k}^{\mu\nu}) = 0 
\label{DT0}
\eeq
must hold on shell.
Note that a similar condition would follow as well using a vierbein formalism,
but with energy-momenta tensors that are obtained using vierbein variations.  

With explicit symmetry breaking,
the four equations in \rf{DT0} do not automatically hold when the dynamical
matter and metric fields are put on shell,
since the background coefficient $\bar k_{\mu\nu\cdots}$ does not
satisfy Euler-Lagrange equations.
Thus, the consistency of the theory depends on whether or not the four additional modes
that arise as a result of the symmetry breaking appear in such a 
way that allows \rf{DT0} to hold.

To examine the role of these extra modes,
a St\"uckelberg approach can be used.
As shown in \rf{Stuck1} and \rf{Stuckmodes},
this gives the extra modes the form of four NG excitations denoted as $\pi^\mu$.
These NG excitations are the same as those that occur in the minimal SME
with spontaneous breaking as shown in \rf{diffNGmodes}.
However, the massive Higgs-like modes in \rf{diffNGmodes} do not
occur when the symmetry breaking is explicit.

Note that the contributions to the action in \rf{explL} with explicit breaking
are separated and organized differently than those in \rf{Sparts} and \rf{LLV}, 
where the breaking is spontaneous.
This is largely due to the absence of the massive Higgs-like excitations
in the case of explicit breaking.
For example, the term ${\cal L}_{\rm R,\bar k}$ in \rf{explL} contains the terms
given in ${\cal L}_{\rm LV}^{\rm (grav)}$ in \rf{llvmsme},
when $\bar k_{\mu\nu\cdots}$ in ${\cal L}_{\rm LV}^{\rm (grav)}$ 
is replaced by the fixed backgrounds
$\bar u$, $\bar s^\mn$ and $\bar t^{\ka\la\mu\nu}$,
and the excitations are limited to just the NG modes.
Similarly, the terms ${\cal L}_{\vp,\bar k}$ in \rf{explL} can overlap
with  ${\cal L}_{\rm LV}^{\rm (matter-grav)}$ in \rf{Sparts} 
when the massive excitations
in  ${\cal L}_{\rm LV}^{\rm (matter-grav)}$ are excluded.
However, terms of the form ${\cal L}_{\rm g,\bar k}$ in \rf{explL} 
are not generally separated out in \rf{Sparts} and \rf{LLV},
though they might emerge from ${\cal L}_{\rm SSB}^{\prime}$ 
in a limit where the excitations decouple.

\subsubsection{Pure-gravity post-Newtonian limit}

In the post-Newtonian limit of the SME
a linearized approach is used
where the NG and massive modes can emerge from both 
${\cal L}_{\rm LV}^{\rm (grav)}$ and ${\cal L}_{\rm SSB}^{\prime}$.
The form of these excitations and their known symmetry properties
allow a systematic method to be applied,
where the NG and massive modes are eliminated in 
terms of the gravitational excitations
\cite{qbak06}.
This permits an expansion in terms of 
gravitational potentials,
and the result is a useful post-Newtonian framework
for investigating Lorentz violation.

Since the SME with explicit breaking has the same NG excitations
as the theory with spontaneous breaking,
it is possible for them to play similar roles in both cases.
This suggests that unless the massive Higgs-like modes have
an essential role in consistently developing the post-Newtonian limit of the SME,
the same systematic approach should work with explicit breaking
as it does in the case of spontaneous breaking.

The main obstacle that has to be overcome to maintain consistency 
with explicit breaking is the requirement of 
covariant energy-momentum conservation in \rf{DT0}, 
which must hold despite the fact that the background $\bar k_{\mu\nu\cdots}$
is nondynamical and does not have Euler-Lagrange equations.
With explicit breaking, it is the four NG modes that must provide solutions 
that allow the four equations in \rf{DT0} to hold,
and in principle the counting of modes suggests this is possible.
However,
in situations where the appearance of the NG modes is suppressed or limited,
it might not then be possible for these modes to provide the needed solutions.
In that case, a useful post-Newtonian limit might not exist. 

The possibility of developing a pure-gravity post-Newtonian limit of the minimal
SME with explicit breaking was examined in \cite{rb15a}.
A pure-gravity sector consisting of an Einstein-Hilbert term and a term
of the form ${\cal L}_{\rm R,\bar k}$ was considered,
where the latter allows couplings between the metric,
the curvature tensor, and the background $\bar k_{\mu\nu\cdots}$.
The NG modes $\pi^\mu$ enter through the substitution \rf{Stuck1} 
and the expansion \rf{Stuckmodes},
and since they appear nonlinearly in 
$T_{\rm R,\bar k}^{\mu\nu}$,
solutions ensuring that $D_\mu T_{\rm R,\bar k}^{\mu\nu}  = 0$ 
in general can exist.

However, in a linearized limit,
the NG modes are suppressed,
which then stands in the way of obtaining a useful post-Newtonian expansion.
This is because the linearized curvature tensor, $R^{\rm (linear)}_{\ka\la\mu\nu}$, 
is invariant under infinitesimal diffeomorphism transformations,
which take the form 
$h_{\mu\nu} \rightarrow h_{\mu\nu} + \partial_\mu \xi_\nu + \partial_\nu \xi_\mu$
in the linearized theory.
Therefore, a term of the form $\bar k^{\ka\la\mu\nu} R^{\rm (linear)}_{\ka\la\mu\nu}$ 
does not contain any of the NG degrees of freedom,
which have the form of virtual diffeomorphisms.
Attempting to use a St\"uckelberg approach does not work either,
since terms $\sim (\partial_\mu \pi^\al) R^{\rm (linear)}_{\al\la\mu\nu}$
are of higher order and must be dropped at the linearized level. 

The result is that the equation $D_\mu T_{\rm R,\bar k}^{\mu\nu}  = 0$ 
in linearized form has no dependence on the NG modes.
Therefore, it is impossible for the NG modes to take values that set $D_\mu T_{\rm R,\bar k}^{\mu\nu}  = 0$.
Since the SME with explicit breaking has no massive Higgs-like excitations,
these excitations cannot play a role like they can with spontaneous breaking.
Instead, the curvature tensor itself must take restricted values to make the 
equation $D_\mu T_{\rm R,\bar k}^{\mu\nu}  = 0$ hold.
For example,
with constant values of $\bar k_{\mu\nu\cdots}$,
partial spacetime derivatives of the curvature tensor are forced to vanish,
which severely limits the geometry of the spacetime.
Even with nonconstant backgrounds $\bar k_{\mu\nu\cdots}$,
severe limitations need to be imposed on the curvature tensor,
which does not allow a useful post-Newtonian limit to be found
\cite{rb15a}.

\subsubsection{Matter-gravity Lorentz-breaking interactions}

When Lorentz-violating matter-gravity couplings are included in the SME,
the NG modes again play a crucial role in developing a
consistent methodology that can be used to identify
observable signals of spacetime symmetry breaking.
Interestingly,
in the context of the SME based on spontaneous breaking,
the approach used in \cite{akjt} makes the assumption that the massive Higgs-like 
excitations are either frozen out or have negligible excitations.
Thus, it is only the NG excitations that are considered
in matter-gravity interactions.
As described in \cite{akjt},
a perturbative treatment can be developed, 
using known symmetry properties, 
which allow the NG modes to be eliminated in terms 
of the gravitational excitations and background SME coefficients.
This methodology allows the dominant signals of spacetime
symmetry to be extracted regardless of the details of the underlying theory.

The reason this approach works despite freezing out the massive modes is
because additional interaction terms involving the background,
metric, and matter terms provide additional couplings to the NG modes.
This allows the NG modes to satisfy equations maintaining the consistency 
of the theory without the need of the massive Higgs-like modes.
It also includes higher-order contributions in a perturbative treatment,
as opposed to restricting the excitations to just the linearized level.

Since this approach used to analyze matter-gravity couplings 
drops the massive Higgs-like excitations,
this same methodology should still apply when the breaking is explicit,
where such excitations do not exist.
The NG modes that occur in a St\"uckelberg approach
with explicit breaking,
can then play the same role as the NG modes with spontaneous breaking,
since both sets have the same mathematical form.

With gravity-matter couplings included,
the SME with explicit breaking has additional Lagrangian terms
besides just ${\cal L}_{\rm R,\bar k}$.
For example,
both ${\cal L}_{\vp,\bar k}$ and ${\cal L}_{\rm g,\bar k}$
can be included in \rf{explL}.
With all three of these terms included,
the consistency conditions in \rf{DT0} involve three 
energy-momentum contributions, 
$T_{\rm R,\bar k}^{\mu\nu}$, $T_{\rm g,\bar k}^{\mu\nu}$, 
and $T_{\vp,\bar k}^{\mu\nu}$.
Thus, even if the NG modes drop out of 
the first term, $D_\mu T_{\rm R,\bar k}^{\mu\nu}$,
in a linearized treatment,
they will still in general appear in the other terms in \rf{DT0}.
In this way, the NG modes can provide solutions without
having to put restrictions on the curvature tensor.

\section{Applications}

As examples, two gravitational theories with explicit
spacetime symmetry breaking are examined in this section.
Both have additional Lagrangian terms matching one or more
of the types of terms included in \rf{explL}.
In ghost-free massive gravity,
potential terms of the form ${\cal L}_{\rm g,\bar k}$ are included
as part of the action,
and matter interactions of the form ${\cal L}_{\vp,\bar k}$
can be considered as well.
In an infrared limit of Ho\v rava gravity using a covariant formulation
that allows matter-gravity interactions,
terms of the form ${\cal L}_{\vp,\bar k}$ can appear in the matter sector.
In both of these examples,
the SME can be used to investigate effects of the explicit 
local Lorentz and diffeomorphism violation that occurs 
in these theories.

\subsection{Massive gravity}

For many years,
attempts to construct a nonlinear gravitational theory
with a massive graviton,
which generalizes the linear Fierz-Pauli theory and agrees
with GR in the massless limit,
remained hindered by the presence of a ghost mode
known as the Boulware-Deser (BD) ghost
\cite{BD72}.
However, more recently, 
the models found by de Rham, Gabadadze, and Tolley (dRGT), 
which contain a particular type of nonlinear interaction involving the metric, 
have been shown to be ghost free
\cite{dRGT,HR}.

A key feature of dRGT massive gravity is that a symmetric background tensor,
denoted here as $\bar f_{\mu\nu}$, must be coupled with the metric
in an interaction potential in order to create mass terms for the metric.
This background is a nondynamical tensor with preferred directions,
and its appearance in the dRGT Lagrangian explicitly breaks diffeomorphism invariance.
In the original versions of dRGT massive gravity,
the background was assumed to be Minkowski,
with $\bar f_{\mu\nu} = \et_{\mu\nu}$.
However,
it was subsequently found that more general backgrounds $\bar f_{\mu\nu}$ can be used,
which need not have constant components.

The dRGT action can be divided into a gravitational sector and a matter sector,
\beq
S_{\rm dRGT} =  S_{\rm grav} + S_{\rm matter} ,
\label{SdRGT}
\eeq
and it can be used to describe massive gravity at
the level of effective field theory in 
either a metric or vierbein formalism.

\subsubsection{Gravity sector}

In a metric description,
the action in the gravity sector has the form
\beq
S_{\rm grav} = \fr 1 {16 \pi G} \int d^4x \sqrt{-g} ( R - \fr {\mu^2} 4 {\cal U}({\mathbb X}) ) ,
\label{SdRGTgrav}
\eeq
where $\mu$ is the graviton mass and
${\cal U}({\mathbb X})$ is a potential formed in terms of square roots ${\mathbb X}^\mu_{\pt{\mu}\nu}$
defined as
\beq
{\mathbb X}^\mu_{\pt{\mu}\nu} = \sqrt{g^{\mu\al} \bar f_{\al\nu}} = \left( \sqrt{g^{-1} \bar f} \right)^\mu_{\pt{\mu}\nu} .
\label{gasqrt1}
\eeq
Effectively, these square roots are matrices that obey
${\mathbb X}^\mu_{\pt{\mu}\al} {\mathbb X}^\al_{\pt{\mu}\nu} = g^{\mu\al} \bar f_{\al\nu}$.
However, their existence is not guaranteed
\cite{dmz},
and often they are obtained in the context of a specific model
that provides ansatz forms for $\bar f_{\mu\nu}$ and the metric $g_{\mu\nu}$.

The potential ${\cal U}({\mathbb X})$ in \rf{SdRGTgrav} is given as,
\beq
{\cal U}({\mathbb X})=   \sum_{n=0}^4 \be_n e_n ({\mathbb X})  .
\label{SdRGTU}
\eeq
It consists of a sum of elementary symmetric polynomials $e_n ({\mathbb X})$
formed from traces of products of ${\mathbb X}^\mu_{\pt{\mu}\nu}$,
with dimensionless couplings $\be_n$ of order one.  
With such a form for ${\cal U}({\mathbb X})$,
the Boulware-Deser ghost does not appear.

Alternatively, in a vierbein description,
a background vierbein $\bar v_{\mu}^{\pt{\mu}a}$ is introduced,
which obeys
\beq
\bar f_{\mu\nu} = \bar v_{\mu}^{\pt{\mu}a} \bar v_{\nu}^{\pt{\mu}b} \et_{ab} .
\label{fvv}
\eeq
When $\bar v_{\mu}^{\pt{\mu}a}$ appears in the Lagrangian,
it explicitly breaks both local Lorentz invariance and diffeomorphisms.
The potential ${\cal U}$ can be defined most simply in a vierbein description 
as the sum of all possible wedge products that can be formed using 
$\vb \mu a$ and $\bar v_{\mu}^{\pt{\mu}a}$.
However, an equivalent expression for ${\cal U}$ can be found
that again has the form of a sum of elementary symmetric polynomials.
In this case, the sums are formed from products and traces of matrices defined as
\beq
\ga^\mu_{\pt{\mu}\nu} = \ivb \mu a \bar v_\nu^{\pt{\nu}a} .
\label{gasqrt}
\eeq
If the vierbein obeys a symmetry condition,
\beq
\ivb \mu a  \bar v_{\mu b} = \ivb \mu b \bar v_{\mu a} ,
\label{symv}
\eeq
then the metric and vierbein descriptions in the absence of matter 
can be shown to be equivalent,
and a solution for ${\mathbb X}^\mu_{\pt{\mu}\nu}$ exists, 
where it equals $\ga^\mu_{\pt{\mu}\nu}$
\cite{hr12}.

Notice that the potential ${\cal U}({\mathbb X})$ in \rf{SdRGTgrav} has the form of an explicit-breaking term 
${\cal L}_{\rm g,\bar k}$ in \rf{explL}, 
when the generic background $\bar k_{\mu\nu\cdots}$
is replaced by $\bar f_{\mu\nu}$.
However, there are important differences between these terms as well.
One difference is that the dependence on $\bar f_{\mu\nu}$ cannot be 
clearly separated out in the potential in \rf{SdRGTgrav},
while in the SME, terms such as ${\cal L}_{\rm g,\bar k}$ would have
a well defined dependence on the background coefficient.
Instead, it is the square root ${\mathbb X}^\mu_{\pt{\mu}\nu}$ that appears 
explicitly in the potential ${\cal U}({\mathbb X})$,
and it has unusual properties.
For example,
while $\bar f_{\mu\nu}$ is a fixed nondynamical background,
${\mathbb X}^\mu_{\pt{\mu}\nu}$ has a hybrid form as the square root
of the dynamical metric contracted with $\bar f_{\mu\nu}$.
Since the background $\bar f_{\mu\nu}$ remains fixed under diffeomorphisms,
${\mathbb X}^\mu_{\pt{\mu}\nu}$ must transform anomalously.
If instead a vierbein description is used,
then there is a clear separation between the dynamical vierbein
$\vb \mu a$ and the background vierbein $\bar v_{\mu}^{\pt{\mu}a}$.
Nonetheless,
their product $\ga^\mu_{\pt{\mu}\nu}$ in \rf{gasqrt} also transforms anomalously
under both diffeomorphisms and local Lorentz transformations.

\subsubsection{Matter sector}

When quantum corrections are taken into account,
it is expected that matter fields in dRGT massive gravity can have couplings to
both the metric $g_{\mu\nu}$ and the background $\bar f_{\mu\nu}$.
In \cite{matterdRGT},
the form that these couplings can take as a result of one-loop interactions was explored, 
and the requirement that they do not introduce the BD ghost was imposed.
The result is that matter fields can couple with an effective metric,
$g^{\rm (eff)}_{\mu\nu}$, which is formed out of both the metric and the background field.
At the same time, the gravity sector remains unchanged,
and the curvature in the Einstein-Hilbert term is constructed using only the physical metric $g_{\mu\nu}$.
The interactions with the background in the pure-gravity sector 
continue to occur only through the potential ${\cal U}({\mathbb X})$,
which is why the ghost does not appear.
Matter couplings were also explored classically,
and a similar effective metric was found by requiring that the weak equivalence principle must hold
while not allowing the BD ghost to appear
\cite{NM15}.
The form of the effective metric that was found in both cases is
\beq
g^{\rm (eff)}_{\mu\nu} = \al^2 g_{\mu\nu} + 2 \al \be g_{\mu\si} {\mathbb X}^\si_{\pt{\mu}\nu} + \be^2 \bar f_{\mu\nu} ,
\label{geff}
\eeq
where $\al$ and $\be$ are constant coupling parameters.
Using instead a vierbein description,
the corresponding form of the effective vierbein that couples to matter is 
\beq 
{e^{\rm (eff)}_{\,\,\,\,\,\,\,\,\, \mu}}^{a} = \al \vb \mu a + \be \bar v_\mu^{\pt{\mu}a} .
\label{eeff}
\eeq

If the symmetry condition in \rf{symv} holds,
the effective metric can be written in terms of $\ga^\mu_{\pt{\mu}\nu}$ as
\beq
g^{\rm (eff)}_{\mu\nu} = \al^2 g_{\mu\nu} + 2 \al \be \ga_{\mu\nu} + \be^2 \bar f_{\mu\nu} ,
\label{geff2}
\eeq
where $\ga_{\mu\nu} = g_{\mu\si} \ga^\si_{\pt{\mu}\nu}$.
Note that with lower indices, the square root matrix is symmetric, 
obeying $\ga_{\mu\nu} = \ga_{\nu\mu}$.

In terms of the effective metric, 
the action in the matter sector has the form
\beq
S_{\rm matter} = \int d^4x \, \sqrt{-g_{\rm (eff)}} \, {\cal L}_{\rm matter} (g^{\rm (eff)}_{\mu\nu}, \vp_\si)  ,
\label{SdRGTmattermetric}
\eeq
or if fermions are included it is given as
\beq
S_{\rm matter} = \int d^4x \, e_{\rm (eff)} \, {\cal L}_{\rm matter} ({e^{\rm (eff)}_{\,\,\,\,\,\,\,\, \mu}}^{a}, \vp_\si,\ps) .
\label{SdRGTmattervierbein}
\eeq

For $\be \ne 0$, matter interactions with $g^{\rm (eff)}_{\mu\nu}$ break local Lorentz symmetry
and diffeomorphisms,
because the matter fields can interact directly with the background field.
Since Lorentz breaking is known to be small,
it is reasonable to assume $\al \simeq 1$ while $\be \ll 1$.
Thus, to first order in $\be$,
the effective metric has the form
\beq
g^{\rm (eff)}_{\mu\nu} \simeq g_{\mu\nu} + 2 \be \ga_{\mu\nu} ,
\label{geff3}
\eeq
while the effective vierbein is
\beq
{e^{\rm (eff)}_{\,\,\,\,\,\,\,\,\, \mu}}^{a} \simeq  \vb \mu a  + \be \vb \al a \ga^\al_{\pt{\al}\mu}  ,
\label{eeffexpans}
\eeq
and their inverses are given approximately as
\bea
g^{{\rm (eff)} \mu\nu} \simeq g^{\mu\nu} - 2 \be \ga^{\mu\nu} ,
\label{invgeff}
\\
e^{{\rm (eff)} \mu}_{\quad\quad a} \simeq  \ivb \mu a  - \be \ga^\mu_{\pt{\al}\al} \ivb \al a  .
\label{inveeffexpans}
\eea
The matter terms in \rf{SdRGTmattermetric} or \rf{SdRGTmattervierbein} 
can be mapped into the SME by expanding the effective 
vierbein or metric and by matching the interactions involving the background 
fields with appropriate SME coefficients.
Alternatively, field redefinitions can be used to change the effective metric
back to the physical metric,
which also results in the appearance of SME coefficients.

As a specific example, consider matter-gravity interactions in massive gravity 
involving photons and a fermion of mass $m$.
The action in this case is given as
\bea
S_{\rm dRGT} = \fr 1 {16 \pi G} \int d^4x \sqrt{-g} ( R - \fr {\mu^2} 4 {\cal U}({\mathbb X}) )
\quad\quad\quad
\nonumber \\
+ \int d^4x \, \sqrt{-g^{\rm (eff)}} \left(  - \fr 1 4 F_{\ka\la} \, g^{{\rm (eff)} \ka\mu} g^{{\rm (eff)} \la\nu}F_{\mu\nu} \right)
\nonumber \\
+ \int d^4x \, e^{\rm (eff)} [ i  {e^{\rm (eff)}}^\mu_{\,\,\, a}  \bar \ps \ga^a D^{\rm (eff)}_\mu \ps 
- m \bar \ps \ps ] .
\label{SdRGT2}
\eea
Here, the pure gravity sector involves the physical metric $g_{\mu\nu}$,
and the curvature is defined in terms of $g_{\mu\nu}$,
while the matter interactions involve the effective metric,
and the covariant derivative $D^{\rm (eff)}_\mu$ is defined
using $g^{\rm (eff)}_{\mu\nu}$.

Field redefinitions with forms similar to those in \rf{tildeg} and \rf{tildee}
can be used on $g^{{\rm (eff)} \mu\nu}$ 
and $e^{{\rm (eff)}\mu}_{\quad\quad a}$,
yielding new effective fields $\tilde g^{{\rm (eff)} \mu\nu}$ and $\tilde e^{{\rm (eff)}\mu}_{\quad\quad a}$
defined by
\beq
g^{{\rm (eff)} \mu\nu} = (1+u) \, \tilde g^{{\rm (eff)} \mu\nu} + s^{\mu\nu} ,
\label{gefftilde}
\eeq
\beq
e^{{\rm (eff)}\mu}_{\quad\quad a} = (1 + \half u) \, \tilde e^{{\rm (eff)}\mu}_{\quad\quad a}
+ \half \tilde e^{{\rm (eff)}}_{\quad \si a} \, s^{\mu\si} .
\label{eefftilde}
\eeq
Using the relations in \rf{Maxwelltilde} and \rf{sckF2},
the matter sector terms in $S_{\rm dRGT}$ can be rewritten in terms
of $\tilde g^{{\rm (eff)} \mu\nu}$ and $\tilde e^{{\rm (eff)}\mu}_{\quad\quad a}$,
which introduces the SME coefficients in \rf{kFdef} and \rf{calbedef}.
Then by choosing specific values for the SME coefficients in terms of $\be$ and $\ga_{\mu\nu}$
given as
\beq
u = - \half \be \ga^\si_\si ,
\label{umassive}
\eeq
\beq
s^{\mu\nu} = - 2 \be (\ga^{\mu\nu} - \fr 1 4 \ga^\si_\si g^{\mu\nu} ) ,
\label{smassive}
\eeq
the redefined effective metric and vierbein are such that they
reduce to the physical metric and vierbein 
at leading order in the SME coefficients:
\beq
\tilde g^{{\rm (eff)} \mu\nu} \simeq g^{\mu\nu} ,
\label{geffgphys}
\eeq
\beq
\tilde e^{{\rm (eff)}\mu}_{\quad\quad a} \simeq e^{\mu}_{\pt{\mu}a} .
\label{eeffgphys}
\eeq
In this way,
the massive gravity action $S_{\rm dRGT} $
is expressed entirely in terms of the physical metric
and the redefined fermion field $\ch$,
but with additional interactions with SME coefficients.
The result is is 
\bea
S_{\rm dRGT} \simeq \fr 1 {16 \pi G} \int d^4x \sqrt{-g} ( R - \fr {\mu^2} 4 {\cal U}({\mathbb X}) )
\quad\quad\quad
\nonumber \\
+ \int d^4x \, \sqrt{-g} \left(  - \fr 1 4 F_{\ka\la} (g^{\ka\mu} g^{\la\nu} + (k_F)^{\ka\la\mu\nu}) F_{\mu\nu}  \right)
\nonumber \\
+ \int d^4x \, e [ i  e^\mu_{\,\,\, a}  \bar \ch ( \ga^a - c_{\al\be} \uvb \be a  \ivb \al b \ga^b) D_\mu \ch 
- m \bar \ch \ch ] .
\label{SdRGT3}
\eea
As a consequence of the field redefinitions,
the Lorentz-violating couplings to the effective metric $g^{\rm (eff)}_{\mu\nu}$ have been replaced
by Lorentz-violating terms involving the SME coefficients $(k_F)^{\al}_{\pt{\al}\mu\al\nu}$ and $c_{\mu\nu}$.
Notice that because the field redefinitions were performed only in the matter sector,
there are no independent gravity sector coefficients $u$ or $s^{\mu\nu}$ in \rf{SdRGT3}.
It is for this reason that the spacetime symmetry breaking is physical as long as $\be \ne 0$
in the definitions of $(k_F)^{\al}_{\pt{\al}\mu\al\nu}$ and $c_{\mu\nu}$ in
\rf{umassive} and \rf{smassive},
since it is not possible to perform further field redefinitions to remove
$(k_F)^{\al}_{\pt{\al}\mu\al\nu}$ and $c_{\mu\nu}$ without generating new 
independent coefficients $u$ and $s^{\mu\nu}$ in the gravity sector.

\subsubsection{Phenomenology}

While the mass potential ${\cal U}({\mathbb X})$ term in \rf{SdRGT3},
which includes a factor of the graviton mass $\mu$ squared,
is essential in giving massive gravity extra degrees of freedom
in the metric while avoiding the ghost mode,
and is important in cosmology and gravitational radiation,
it has a negligible effect in matter-gravity tests performed in laboratories
on Earth or in space experiments on solar system scales.  
This is because the graviton mass is experimentally bounded to
extremely small values of order $\mu \lsim 10^{-29}$ eV
\cite{nieto10}.

In the context of matter-gravity tests,
the gravitational interaction can be modeled using a
post-Newtonian limit and a perturbative treatment in the context of the SME,
where the effects of $\mu$ can be ignored at leading order.
Instead, the effects of the interactions with the background field
can be probed,
which depend on the parameter $\be$.
Since the metric $g_{\mu\nu}$ and the background $\bar f_{\mu\nu}$ are 
typically approximated at lowest order as Minkowski backgrounds,
it follows that the contributions of $\ga^{\mu\nu}$ in the SME coefficients
$(k_F)^{\al}_{\pt{\al}\mu\al\nu}$ and $c_{\mu\nu}$ will be of order one.
Hence,
the extent of the spacetime symmetry breaking is determined primarily by $\be$
in \rf{geff3} and \rf{eeffexpans},
and it is therefore this parameter that can be used as a phenomenological measure of 
potential Lorentz violation in the matter sector of massive gravity.  

To investigate the phenomenology of matter-gravity couplings in massive gravity,
experiments with sensitivity to two sets of SME coefficients must be analyzed.
While an $s^{\mu\nu}$ term is missing in \rf{SdRGT3},
a field redefinition of the metric $g^{\mu\nu}$ in the full action $S_{\rm dRGT}$
would change $(k_F)^{\al}_{\pt{\al}\mu\al\nu}$ and $c_{\mu\nu}$ while
also introducing a term $s^{\mu\nu}$ in the gravity sector.
Hence a suitable experiment that can place bounds on the Lorentz-breaking
matter-gravity interactions in massive gravity are the matter-interferometry experiments
that have sensitivity to spacetime symmetry breaking in both the gravity 
and electromagnetic sectors of the SME \cite{atomint}.

The experiments in \cite{atomint} place bounds at the level of $10^{-9}$ on
combinations of $s^{\mu\nu}$ and the coefficients that contribute to
the symmetric trace $(k_F)^{\al\mu\pt{\al}\nu}_{\pt{\al\mu}\al}$.
With the assumption that $|\ga^{\mu\nu}| \simeq 1$,
this gives a bound of
\beq
\be \lsim 10^{-9}
\label{bebound}
\eeq
on the coupling parameter that determines the strength of
the Lorentz-violating interactions between matter 
and the fixed background field in massive gravity.

\subsection{Ho\v rava gravity}

In Ho\v rava gravity, 
diffeomorphism invariance is broken explicitly by the presence of a preferred
foliation of spacetime
\cite{ph09}.
The preferred foliation introduces a physical distinction between time and space,
which can be labeled using coordinates $t$ and $x^i$,
with $i = 1,2,3$,
where constant values of $t$ distinguish the preferred spatial foliations,
and $x^i$ labels the spatial points.
In particular, anisotropic scaling is introduced between $t$ and $x^i$,
which permits higher-dimensional terms to be added in the action involving only spatial derivatives
acting on the metric or matter fields,
while maintaining terms with just two time derivatives.
This in turn allows for the construction of gravity models with power-counting renormalizability.

The preferred foliation breaks the full diffeomorphism group to a subgroup,
consisting of three-dimensional spatial diffeomorphisms and time reparametrizations,
\bea
x^i &\rightarrow& x^i + \xi^i (t,x^j) ,
\label{xdiffF}
\\
t &\rightarrow& t+ \xi^0 (t) .
\label{tdiffF}
\eea
These transformations are called foliation-preserving diffeomorphisms.

The action in Ho\v rava gravity can be divided into three sectors,
\beq
S_{\rm Horava} = S_{\rm K} + S_{\rm V} + S_{\rm M} ,
\label{Shorava}
\eeq
consisting of kinetic (K) and potential (V) terms in the
gravity sector as well as a matter sector (M).
The usual four-dimensional diffeomorphism transformations in GR are explicitly broken 
when they are applied to the action $S_{\rm Horava}$.
Instead, it is the foliation-preserving diffeomorphisms
that are the fundamental spacetime symmetry in Ho\v rava gravity.  

The low-energy or infrared (IR) limit of Ho\v rava gravity
must approach GR and the SM if it is to be phenomenologically viable.
This requires that coupling constants associated with the
spacetime symmetry breaking must have limits consistent
with GR and the SM in the IR limit.
To make connections with the SME,
the IR limit of Ho\v rava gravity must be expressed in
a covariant form,
and correspondence with appropriate SME coefficients
must be identified.
The remainder of this section will only consider
Ho\v rava gravity in the IR limit, 
and it will investigate how the SME can be used to examine 
spacetime symmetry breaking in matter-gravity couplings
in this context.

\subsubsection{Gravity sector}

To construct the gravity sector of the action in
the IR limit of Ho\v rava gravity,
the four-dimensional metric $g_{\mu\nu}$ is
replaced by the ADM variables $(N,N^i,g_{ij})$
consisting of the lapse, the shift, and the three-dimensional spatial metric.
These become the fundamental fields of the theory
and are used to define the three-dimensional Ricci tensor $R^{\rm (3)}_{ij}$,
the extrinsic curvature $K_{ij}$, and covariant derivatives $D_j$.
Lagrangian terms can then be constructed by contracting these field operators to form
scalars under spatial diffeomorphisms \rf{xdiffF} and time reparametrizations \rf{tdiffF}.
The spacetime integrals in the action $S_{\rm Horava}$ also include
factors of $\sqrt{g^{\rm (3)}} N d^3x \, dt$,
where $g^{\rm (3)}$ is the determinant of $g_{ij}$.

Time derivatives of $g_{ij}$ are kept at second order
to prevent the appearance of ghosts,
and they enter the Lagrangian through the extrinsic curvature,
which is defined as
\beq
K_{ij} = \fr 1 {2N}(- \dot g_{ij} + D_i N_j + D_j N_i ) .
\label{extrins}
\eeq
The kinetic term ${\cal L}_{\rm K}$ is defined in terms of
the extrinsic curvatures as
\beq
{\cal L}_{\rm K} = \fr 1 {16 \pi G}  (K_{ij} K^{ij} - \la_g K^2) ,
\label{LK}
\eeq
where $K = g^{ij} K_{ij}$, 
and $\la_g$ is a running coupling constant.
Note that the two terms in \rf{LK} are each independently scalars under
foliation-preserving diffeomorphisms,
and $\la_g$ gives the relative weighting between these terms.

The potential term ${\cal L}_{\rm V}$ consists of contractions 
of spatial components,
which in most versions of Ho\v rava gravity includes terms with 
up to dimension-six operators.
It is the inclusion of the higher-dimensional terms involving spatial derivatives that makes
Ho\v rava gravity power-counting renormalizable in the high-energy limit.
However, the leading-order terms at low energy
are the three-dimensional curvature scalar
and a cosmological constant term.
The potential term then has the form
\beq
{\cal L}_{\rm V} = \fr 1 {16 \pi G} (R^{\rm (3)} - 2 \La) + \cdots ,
\label{LV}
\eeq
when the higher-order terms are not included.
Note that different versions of Ho\v rava gravity include different combinations 
of fields in the higher-order terms,
and in certain cases additional internal symmetries are included.
However, in the context of this discussion the key element is the explicit 
breaking of timelike diffeomorphisms,
which is a common feature of all types of Ho\v rava gravity models,
and the specific form of the higher-dimensional operators is not important.
In particular, in the IR limit, the higher-order terms all become small in
comparison to the terms shown in \rf{LV}.

For Ho\v rava gravity to match with GR in the IR limit,
the running coupling $\la_g$ must approach 1.
With $\la_g = 1$, the terms ${\cal L}_{\rm K}$ and ${\cal L}_{\rm K}$
reproduce the usual Einstein-Hilbert and $\La$ terms in GR
in ADM formalism.
It is assumed that when these terms combine in the IR limit,
the four-dimensional metric and the reconstructed Einstein-Hilbert term
recover their usual transformation properties.
In particular, for these terms the full diffeomorphism invariance is restored when $\la_g = 1$.
However, with small residual values of $(1- \la_g)$ at low-energy scales,
there is still some spacetime symmetry breaking,
which results in an additional symmetry-breaking term in the IR limit.
With these assumptions,
the action of the gravity sector of Ho\v rava gravity in the IR limit can be written as
\bea
S_{\rm K} + S_{\rm V} \simeq \int \sqrt{-g} \, d^4 x \fr 1 {16 \pi G} (R - 2 \La)  
\nonumber \\
+ (1-\la_g) \fr 1 {16 \pi G} \int \sqrt{g^{\rm(3)}} N d^3 x dt \, K^2  .
\label{ShoravaIR}
\eea
Here, it can be seen that the parameter $(1-\la_g) \ll 1$ becomes the primary measure of 
the spacetime symmetry breaking in the gravity sector.

The residual symmetry-breaking term in \rf{ShoravaIR} can be put in covariant form
by introducing a timelike unit vector $n_\mu$.
In terms of the coordinates $(t,x^i)$,
it is given by
 $n_\mu = (-N,0)$ and
$n^\mu = (\fr 1 N, - \fr {N^i} N)$, 
which obey $n_\mu n^\mu = -1$.
Using these, 
the four-dimensional metric $g^{\mu\nu}$ can be given in terms of the three-dimensional 
spatial metric $g^{ij}$ and the normal vectors $n^\mu$ as
\beq
g^{\mu\nu} = g^{ij} \de^\mu_i \de^\nu_j - n^\mu n^\nu ,
\label{g43}
\eeq
The Kronecker delta functions such as $\de^\mu_i$ appear as a result of using the coordinates $(t,x^i)$.
If a foliation-preserving coordinate transformation to new coordinates $x^{\mu^\prime}$ is performed,
the Kronecker delta $\de^\mu_i$ gets transformed into coordinate transformation matrices
\beq
e^{\mu^\prime}_{\pt{\mu}i^\prime} = \fr {\partial x^{\mu^\prime}} {\partial x^\al} \fr {\partial x^j} {\partial x^{i^\prime}} \de^\al_j ,
\label{emui}
\eeq
and alternative expressions using $e^{\mu^\prime}_{\pt{\mu}i^\prime}$ can be obtained.
However, the main results found using these more general matrices 
can also be found in simpler form using $\de^\mu_i$ and coordinates $(t,x^i)$.
For this reason, 
the coordinates $(t,x^i)$ are used in the remainder of this section.

The timelike unit vector $n_\mu$ can also be used to define a projection operator.
First define
\bea
h^{\mu\nu} &=& g^{ij} \de^\mu_i \de^\nu_j 
\nonumber \\
&=& g^{\mu\nu} + n^\mu n^\nu ,
\label{hmunu}
\eea
which obeys $h^{\mu\nu} n_\mu = 0$.
Its mixed form is given as
\beq
h^{\mu}_{\pt{\mu}\nu} = \de^{\mu}_\nu + n^\mu n_\nu ,
\label{hmunu2}
\eeq
which defines a projection operator that can be
used to project tensors in the four-dimensional spacetime
into the three-dimensional spatial foliation while maintaining covariance.

Expressions involving the extrinsic curvature $K_{ij}$ can be defined in terms of $n_\mu$ as well.
For example, in $(t,x^i)$ coordinates,
it can be shown that \cite{poisson}
\beq
K_{ij} = \de^\mu_i \de^\nu_j D_\nu n_\mu .
\label{Kijn}
\eeq
and that the extrinsic curvature is symmetric,
obeying $K_{ij} = K_{ji}$.
Its trace $K = g^{ij} K_{ij}$ in covariant form in the
four-dimensional spacetime is then given as
\bea
K &=& h^{\mu\nu} D_\nu n_\mu 
\nonumber \\
&=& ( g^{\mu\nu} + n^\mu n^\nu) D_\nu n_\mu .
\label{KhDn}
\eea
With this expression,
the gravity sector of Ho\v rava gravity in the IR limit
can then be written in covariant form as
\bea
S_{\rm K} + S_{\rm V} \simeq \int \sqrt{-g} \, d^4 x \fr 1 {16 \pi G} [R - 2 \La
\quad\quad
\nonumber \\
+ (1-\la_g) K^2]  .
\label{covShoravaIR}
\eea

It is important to realize, however, that despite its covariant form,
the action for the gravity sector of Ho\v rava gravity 
still explicitly breaks timelike diffeomorphisms.
This is because the vector $n_\mu$ becomes a background field that cannot transform
under timelike diffeomorphisms when $\xi^0$ depends on $x^j$.
It must remain normal to the preferred foliation.
In this way, contractions with $n_\mu$ are similar to couplings to SME coefficients,
which transform as tensors under general coordinate transformations,
but which remain fixed under diffeomorphisms.
A significant difference from the way the SME coefficients are usually thought of,
however, is that $n_\mu$ is only {\it partially} fixed under diffeomorphisms.
The subgroup consisting of foliation-preserving diffeomorphisms,
defined with $\xi^i (x^j,t)$ and $\xi^0 (t)$,
where the latter only has time dependence,
still transform $n_\mu$ into new physically equivalent normal vectors,
and these transformations are symmetries of the action.
However, the timelike diffeomorphisms with $\xi^0$ depending on position are broken,
since $n_\mu$ remains fixed under such transformations.

Thus, similarly to massive gravity,
where the tensors $\ga_{\mu\nu}$ are not fully fixed under diffeomorphisms and
instead transform anomalously,
the backgrounds $n_\mu$ are only partially fixed under diffeomorphisms
and as a result they too transform anomalously.
In this way, both of these backgrounds differ from how the SME coefficients are usually defined.
Nonetheless, for the symmetries or partial symmetries that are explicitly broken,
the relevant couplings in a theory with explicit breaking can still be matched to
corresponding couplings in the SME,
and any bounds that have been obtained in the SME
can in principle be applied to the theories with explicit breaking.

In parallel with the case of massive gravity,
a St\"uckelberg approach can also be used to describe Ho\v rava gravity.
In the case of Ho\v rava gravity, however,
only one St\"uckelberg scalar is introduced because only one diffeomorphism is broken.
In a St\"uckelberg approach,
the normal vector $n_\mu$ is replaced by $\partial_\mu \Ph$,
where $\Ph (t,x^j)$ is the St\"uckelberg field.
The scalar $\Ph$ is assumed to be dynamical,
which restores the broken timelike diffeomorphism
while at the same time introducing one extra degree of freedom.
The original form of Ho\v rava gravity with explicit breaking can
be obtained again by setting $\Ph = t$.
Thus, in many versions of Ho\v rava gravity,
unless an extra internal symmetry is introduced that can be used to remove $\Ph$,
a primary effect of Ho\v rava gravity is that there is an 
extra scalar degree of freedom in gravitational interactions.

Just as different versions of Ho\v rava gravity have been proposed and explored
in the pure gravity sector
\cite{HGreviews}, 
there are correspondingly different ideas that can be considered for how to couple matter to Ho\v rava gravity
\cite{hmatter}.
The broadest and most general approach, however,
which parallels the way in which the gravity sector is defined,
is to consider a matter action $S_{\rm M}$ where foliation-preserving 
diffeomorphism invariance is the fundamental spacetime symmetry instead
of the full diffeomorphism group.
This is the approach that is examined here.

By explicitly breaking timelike diffeomorphisms,
the time and spatial components of matter fields and their derivatives can be separated
and treated differently similarly to how the ADM fields for the metric are treated differently in the gravity sector.
To avoid ghosts, the usual forms for time derivatives of matter fields,
restricted to second order, can be maintained,
while higher-dimensional terms for the spatial components can be added to the action.
Coupling coefficients can be introduced to give relative weightings between 
these separated terms,
where each term is individually symmetric under foliation-preserving diffeomorphisms.
It is expected that these weighting coefficients can run with energy
and that they must reduce to values consistent with GR and the SM in the IR limit.
There is also no reason to assume that these couplings are the same in different particle sectors
or that they should directly be related to the parameter $\la_g$ in the gravity sector.
Thus, different particle sectors need to be considered independently.

In the absence of a vierbein formalism for Ho\v rava gravity,
which would be needed to consider couplings to fermion fields,
the examples considered here look at the cases of couplings to
massive scalar and massless vector particles,
such as the Higgs boson and the photon.  
The SME is used to investigate the phenomenology of matter-gravity couplings
to these types of particles in the context of Ho\v rava gravity.  

\subsubsection{Scalar Matter Fields}

The simplest case to consider is a scalar matter field $\ph$.
In GR, the usual matter terms for a scalar of mass $m$ interacting with gravity are
\beq
S_{\rm GR, scalar} = \int \sqrt{-g} \, d^4 x (\half g^{\mu\nu} \partial_\mu \ph \, \partial_\nu \ph - \half m^2 \ph^2) .
\label{SGRscalar}
\eeq
Using ADM variables for the metric, the usual four-dimensional kinetic term for the scalar can be rewritten as
\beq
\half \partial_\mu \ph \partial^\mu \ph = - \fr 1 {2N^2} (\dot \ph - N^i \partial_i \ph)^2 + \half g^{ij} \partial_i \ph \, \partial_j \ph ,
\label{scalarADM}
\eeq 
where $\dot \ph = \partial_0 \ph$.
In Ho\v rava gravity the two terms in \rf{scalarADM} are each independently invariant under
foliation-preserving diffeomorphisms,
and therefore they can be given different weightings.
In addition, higher-dimensional operators that are invariant under foliation-preserving diffeomorphisms
can be added to the action,
as long as they do not introduce additional time derivatives
that modify the kinetic term.
However, in the IR limit,
the higher-order couplings will be sub-leading order corrections and
can be ignored here.

Taking different weightings of the two terms in \rf{scalarADM},
the action for a massive scalar in the IR limit of Ho\v rava gravity can be written as
\bea
S_{\rm scalar} \simeq 
\int \sqrt{g} N \, d^3 x \, dt  \left[ c_1^{(\ph)} (- \fr 1 {2N^2} (\dot \ph - N^i \partial_i \ph)^2 ) \right.
\nonumber \\
\left. + c_2^{(\ph)} ( \half g^{ij} \partial_i \ph \, \partial_j \ph) - m^2 \ph^2 \right]
\label{Sph}
\eea
where $c_1^{(\ph)}$ and $c_2^{(\ph)}$ have been introduced as weighting parameters.

It is also possible to use projections of the derivatives,
which gives
\bea
\half g^{ij} \partial_i \ph \partial_j\ph 
&=& \half g^{ij} \de^\mu_i \de^\nu_j \partial_\mu \ph \partial_\nu \ph
\nonumber \\
&=& \half (g^{\mu\nu} + n^\mu n^\nu) \partial_\mu \ph \partial_\nu \ph ,
\label{projph}
\eea
where \rf{g43} has been used to replace the three-dimensional spatial metric $g^{ij}$
with the four-dimensional metric $g^{\mu\nu}$.
By combining \rf{scalarADM} and \rf{projph},
the action $S_{\rm scalar}$ becomes
\bea
S_{\rm scalar} \simeq 
\int \sqrt{-g} \, d^4 x \left[ c_2^{(\ph)} (\half \partial_\mu \ph \partial^\mu \ph ) - m^2 \ph^2  \right.
\nonumber \\
\left. + (c_2^{(\ph)} - c_1^{(\ph)}) ( \half n^\mu n^\nu \partial_\mu \ph \, \partial_\nu \ph) \right]
\label{Sph2}
\eea
Notice that this is now in covariant form.

By rescaling the field $\ph$ and the mass $m$,
the diffeomorphism-invariant term can be put in standard form,
leaving just a relative parameter that multiplies the symmetry-breaking term.
The rescaled field and mass can be relabeled again as $\ph$ and $m$.
These rescalings and relabelings effectively set $c_2^{(\ph)} = 1$ and rename $c_1^{(\ph)}$ as $\la_\ph$,
which gives the final form of the scalar action as
\bea
S_{\rm scalar} \simeq 
\int \sqrt{-g} \, d^4 x \left[ \half g^{\mu\nu} \partial_\mu \ph \, \partial_\nu \ph - \half m^2 \ph^2 \right.
\nonumber \\
\left. + \half (1 - \la_\ph) n^\mu n^\nu \partial_\mu \ph \, \partial_\nu \ph \right].
\label{Sph3}
\eea
In this way, $(1 - \la_\ph$) becomes a measure of the diffeomorphism breaking in the IR limit,
similar to how $(1 - \la_g)$ gives a corresponding measure in the gravity sector.
Also, $\la_\ph$ can run as the energy scale changes
just as $\la_g$ does in the gravity sector.
Agreement with GR and the SM in the IR limit requires $(1 - \la_\ph) \ll 1$.

Notice that the spacetime symmetry breaking in the last term  in \rf{Sph3} 
is due to the coupling $\half ( 1- \la_\ph) n^\nu n^\nu$ acting as a background that explicitly 
breaks timelike diffeomorphisms
while maintaining the foliation-preserving subgroup.
Thus, a connection with an SME coefficient that couples in the same way can be made.
If experimental bounds exist for the corresponding SME coefficient,
they can be applied to the couplings in \rf{Sph3}.
However, since $|n^\mu| \simeq 1$,
the primary result will be that a bound can be placed on
the small coupling $(1 - \la_\ph)$.

The Higgs boson is the only elementary particle in the SM that is a scalar,
and it can be used as a specific example.
In the SME, there is sensitivity to Lorentz violation in the Higgs sector,
and it has been investigated and tested experimentally.
One of the SME coefficients in the Higgs sector is given as $(k_{\ph\ph})^{\mu\nu}$,
which couples the same way as $ \half (1 -  \la_\ph ) n^\nu n^\nu$ in Eq.\ \rf{Sph3}.
Thus, a correspondence can be made,
which gives
\beq
(k_{\ph\ph})^{\mu\nu} = \half (1 - \la_\ph ) n^\nu n^\nu .
\label{kphph}
\eeq

With three unbroken spatial diffeomorphisms,
a gauge can be fixed that sets $N^i = 0$,
which then gives $n^\mu = (\fr 1 N,0)$.
This can be done in any coordinate system,
including Sun-centered celestial equatorial coordinates,
which are used for comparison purposes in the SME.
Note that this procedure using gauge fixing is very 
different from the traditional SME
based on spontaneous spacetime symmetry breaking.
In the traditional case,
the background tensor is fixed under all diffeomorphisms,
and no gauge choices can be made to simplify it.
Special coordinates can always be chosen to
simplify its form;
however, it cannot be assumed that a specific simplified form holds
in any given frame,
such as the Sun-centered celestial equatorial frame.
In the traditional form of the SME,
all of the components of a background tensor must be
assumed to be nonzero in the Sun-centered celestial equatorial coordinate system,
but this is no longer the case with explicit breaking when
only a part of the symmetry is broken by the background field.

As a result of this partial breaking and choice of gauge, 
there is effectively only a purely timelike component of the SME coefficient
$(k_{\ph\ph})^{\mu\nu}$ that is nonzero in the Higgs sector,
and it is given as
\beq
(k_{\ph\ph})^{00} = \fr 1 {2 N^2} (1 - \la_\ph ) .
\label{kph00}
\eeq
Experiments looking to test this type of spacetime symmetry breaking
therefore need to have sensitivity to purely timelike interactions.

While experimental bounds have been obtained on the SME coefficients
$(k_{\ph\ph})^{00}$ in the Higgs sector,
the experiments done to date all assume a Minkowski spacetime,
and ignore gravitational effects.
These tests therefore cannot provide meaningful bounds in the context of Ho\v rava gravity
on the parameter $(1 - \la_\ph )$ for the Higgs.
To obtain a physically meaningful bound,
sensitivity to both gravity and matter is required in order to avoid ambiguities associated 
with the ability to make field redefinitions involving the metric.

\subsubsection{Photons}

The case of a vector particle, such as the photon $\ga$, is considered next.
Using ADM variables for the metric,
the usual four-dimensional Lagrangian term for a massless vector under diffeomorphisms
can be written as
\bea
- \fr 1 4 F^{\mu\nu} F_{\mu\nu} = \fr 1 {2N^2} (g^{ij} - \fr {N^i N^j} {N^2}) F_{0i} F_{0j}
\quad\quad\quad\quad
\nonumber \\
- \fr {N^i} {N^2} (g^{jk} - \fr {N^j N^k} {N^2}) F_{0j} F_{ik}
\quad\quad\quad
\nonumber \\
- \fr 1 4 (g^{ij} - \fr {N^i N^j} {N^2}) (g^{kl} - \fr {N^k N^l} {N^2}) F_{ik} F_{jl} .
\label{max}
\eea
For simplicity, 
a gauge-fixed form of this term is examined,
where the spatial diffeomorphisms are used to set $N^i = 0$.
This reduces the usual Lagrangian to
\beq
- \fr 1 4 F^{\mu\nu} F_{\mu\nu} = \fr 1 {2N^2} g^{ij} F_{0i} F_{0j}
- \fr 1 4 g^{ij} g^{kl}  F_{ik} F_{jl} .
\label{max2}
\eeq

These two terms become the independent terms in gauge-fixed form.
Summing them with weighting parameters $c_1^{(\ga)}$ and $c_2^{(\ga)}$ gives
for the case of a photon field in Ho\v rava gravity
\bea
S_\ga \simeq 
\int \sqrt{g} N \, d^4 x \, dt [c_1^{(\ga)} \fr 1 {2N^2} g^{ij} F_{0i} F_{0j}
\quad
\nonumber \\
- c_2^{(\ga)} \fr 1 4 g^{ij} g^{kl}  F_{ik} F_{jl} ] .
\label{Svec}
\eea
The projections in \rf{g43} can then be used to 
obtain the following two expressions:
\beq
g^{ij} F_{0i} F_{0j} = (g^{\mu\nu} + n^\mu n^\nu) F_{0\mu} F_{0\nu} ,
\label{gijFF}
\eeq
\beq
g^{ij} g^{kl} F_{ik} F_{jl} = (g^{\ka\la} g^{\mu\nu} 
+ 2 g^{\mu\nu} n^\ka n^\la) F_{\ka\mu} F_{\la\nu} .
\label{ggFF}
\eeq
These can be combined with \rf{max2} to rewrite \rf{Svec} in covariant form.
At the same time, the parameters in \rf{Svec} can be redefined as 
$c_1^{(\ga)} = \la_\ga$ and $c_2^{(\ga)} = 1$
so that the four-dimensional kinetic term has its usual form.
The resulting action for a massless vector in the IR limit of Ho\v rava gravity is
\bea
S_\ga \simeq 
\int \sqrt{-g} \, d^4 x (- \fr 1 4 F^{\mu\nu} F_{\mu\nu} 
\quad\quad
\nonumber \\
-  \fr 1 4 (k_F)^{\ka\la\mu\nu} F_{\ka\la} F_{\mu\nu}  ) ,
\label{Svec2}
\eea
where a photon sector SME coefficient $(k_F)^{\ka\la\mu\nu} $ defined as
\bea
(k_F)^{\ka\la\mu\nu} = \half (1 - \la_\ga) [ g^{\ka\mu} n^\la n^\nu 
- g^{\ka\nu} n^\la n^\mu
\quad
\nonumber \\
- g^{\la\mu} n^\ka n^\nu + g^{\la\nu} n^\ka n^\mu ] 
\label{kFHorava}
\eea
has been introduced.
In this context, $n^\nu$ acts as a partially fixed background,
which does not transform under timelike diffeomorphisms
when $\xi^0$ depends on $x^i$.
When the gauge choice with $n^\mu = (\fr 1 N,0)$ is applied,
the SME components $(k_F)^{\ka\la\mu\nu} $ reduce to
\bea
(k_F)^{ijkl} &=& 0 ,
\nonumber \\
(k_F)^{0ijk} &=& 0 ,
\nonumber \\
(k_F)^{0i0j} &=& \fr 1 {2N^2} (1 - \la_\ga) \, g^{ij} .
\label{kvalues}
\eea
Thus, since $|g^{ij}| \simeq 1$,
matter-gravity experiments with sensitivity to the SME coefficients $(k_F)^{0i0j}$
can be used to put bounds on $(1 - \la_\ga)$ in Ho\v rava gravity combined
with electromagnetism.

In this case, experiments with sensitivity to both gravity and the photon sector have been performed.
In particular, the same atom interferometry tests that give bounds on matter-gravity interactions
in massive gravity can also give bounds on possible photon-gravity interactions in Ho\v rava gravity.
By adopting a convention where field redefinitions in the metric are made that eliminate
the gravity-sector $s^{\mu\nu}$ SME coefficients,
this leaves only the sensitivity to $(k_F)^{\ka\la\mu\nu}$ in these matter interferometry tests.
The quantities $\si^{JK}$ that are bounded at the level of $10^{-9}$ in the
Sun-centered celestial equatorial frame
can be applied to the SME coefficients in \rf{kvalues} to give the bound
\beq
|1 - \la_\ga | \lsim 10^{-9}
\label{lphbound} 
\eeq
associated with the spacetime symmetry breaking involving photons
in Ho\v rava gravity.

\section{Summary and Conclusions}

The traditional SME based on the idea of spontaneous spacetime symmetry breaking
is widely used in gravitational, astrophysical, particle, nuclear, solid matter, and atomic experiments aimed 
at testing local Lorentz and diffeomorphism invariance.
When gravity is present,
the fact that the breaking is spontaneous avoids potential inconsistency
between the Bianchi identities, dynamics, and covariant energy-momentum conservation.
Also, with spontaneous breaking,
excitations consisting of NG and massive Higgs-like modes occur,
and knowledge of their behavior allows systematic procedures to be developed
for taking a post-Newtonian limit of the SME and for incorporating matter-gravity
interactions in a consistent manner.

This paper looks at the question of whether the SME can also be used to investigate
gravity theories with explicit spacetime symmetry breaking.
With explicit spacetime symmetry breaking, 
there are nondynamical background fields that appear directly in the action,
and it is the interactions with these backgrounds that cause the symmetry breaking.
At the same time, to be observer independent a gravity theory with explicit breaking
must still be covariant under general coordinate transformations.
The requirement of covariance can be used to derive four mathematical identities that 
must hold in order for the theory to be consistent with the Bianchi identities 
and covariant energy-momentum conservation.
Since four extra degrees of freedom exist in a theory with explicit diffeomorphism breaking,
due to the loss of four gauge freedoms,
these four extra modes can in principle take values that permit the overall consistency
conditions to hold.

It is found using a St\"uckelberg approach that the extra degrees of freedom in
a theory with explicit breaking have the same form as the NG excitations in the
corresponding theory where the symmetry breaking occurs spontaneously.
Thus, many of the procedures and results that follow from having NG modes in the
theory with spontaneous breaking can also be applied when the breaking is explicit.
The main difference is that with explicit breaking,
there are no massive Higgs-like excitations and the background field
remains nondynamical.
In the pure-gravity sector,
the consistency of the theory with explicit breaking therefore
relies completely on the presence of the extra NG modes.
If one or more of these modes is suppressed or decouples,
then the consistency conditions cannot be fulfilled.
An example of when this happens is in the linearized post-Newtonian limit of
the pure-gravity sector of the SME when the symmetry breaking is explicit.
The NG modes decouple in this limit,
and the consistency conditions impose severe constraints on the curvature tensor,
resulting in a theory that is not useful.
However, when matter fields are included,
there are additional interactions that can include the NG modes.
In this case, the same procedures that are used in the SME with spontaneous
breaking carry over and can be used as well when the breaking is explicit.
Thus, the SME is in general suitable for investigating matter-gravity interactions
in theories with explicit breaking.

With gravity,
the role of field redefinitions that can be made involving the metric
is important to consider as well before meaningful physical bounds on
spacetime symmetry breaking can be determined from a specific experiment.
In particular, field redefinitions of the metric can be used to move the
sensitivity to spacetime symmetry breaking from one matter sector to another
or from the gravity sector to a matter sector (or vice versa).
Specifically, any one of the three sets of SME coefficients 
$s^{\mu\nu}$, $c_{\mu\nu}$, or $(k_F)^{\al\mu\pt{\al}\nu}_{\pt{\al\mu}\al}$
can be eliminated from the theory at first order in these coefficients
by making field redefinitions of the metric.
It is for this reason that matter-gravity tests in the SME must have 
sensitivity to at least two independent sets of SME coefficients.

There are a number of different features that occur in the SME
when it is applied to explicit breaking in comparison to the traditional
approach based on spontaneous breaking.
Most notable is that the background fields that cause explicit breaking
are not as clearly defined as they are with spontaneous breaking,
where they are understood as vacuum expectation values.
For example, with explicit breaking,
backgrounds that are hybrids of dynamical and nondynamical fields
or that partially break a spacetime symmetry can appear.
These backgrounds transform anomalously under spacetime
symmetry transformations,
and in some cases they can be partly gauge-fixed using unbroken symmetries.
In making a correspondence with the SME,
the identified SME coefficients might then have only certain components
that are nonzero, 
not just in a special frame but in whatever frame one chooses.
This is clearly a very different feature of explicit breaking in
that it allows time and spatial directions to be physically
distinguished and treated differently.

Two gravity theories with explicit diffeomorphism breaking
serve as examples of how the SME can be used to investigate 
matter-gravity interactions that might occur in these theories.
The first is ghost-free massive gravity with matter interactions that
couple to an effective metric that consists of a linear combination of
the physical metric and a background.
It is shown that the matter terms with couplings to the effective metric can be
replaced by conventional couplings to the physical metric as well
as additional terms that can be matched to the SME.
The second example is Ho\v rava gravity in the IR limit with matter terms 
that have foliation-preserving diffeomorphism
invariance as their fundamental symmetry,
just as this is the fundamental symmetry in the pure-gravity sector.
It is shown that mixtures of these matter terms can be replaced by
a conventional relativistic term plus terms that match those in the SME.
In both examples,
bounds on the diffeomorphism-breaking matter-gravity couplings
are obtained using the SME.
Atom interferometry tests with sensitivity to both gravity and
electromagnetism provide bounds on the order of $10^{-9}$ in
both examples.

\appendix*
\section{Background Vierbeins and the SME}

This appendix illustrates how the background vierbein $\barvb \mu a$ has an
important role in theories with explicit spacetime symmetry breaking.
In particular, it needs to be included in order to go between local
and spacetime frames 
\cite{rbas16}.
Related to this,
it is also shown that the choice of whether to couple matter with a background field 
using local versus spacetime components makes a difference,
and this difference has important consequences concerning consistency 
conditions that must hold with explicit breaking.

For simplicity,
consider a theory with a vector background field, where a vierbein formalism is used.
The background vector has components $\bar k_\mu$ with respect to the spacetime
frame and components $\bar k_a$ with respect to a local Lorentz frame.
Since both sets of these components are fixed under spacetime diffeomorphisms 
and local Lorentz transformations,
they must be connected by a nondynamical background vierbein $\barvb \mu a$,
which is also fixed under these transformations.
The relation between them is 
\beq
\bar k_\mu = \barvb \mu a \bar k_a .
\label{kmuka}
\eeq

If the background vierbein $\barvb \mu a$ is not included in a theory
with explicit breaking,
then actions that couple $\bar k_\mu$ to matter and gravitational fields
are different from actions that couple $\bar k_a$.
To demonstrate this,
consider the following two action terms defined using, respectively, $\bar k_a$ and $\bar k_\mu$ 
to couple to the dynamical gravitational and matter fields:
\beq
S_{\rm 1, \, LV}^{(k_a)} = \int d^4 x \, e \, \bar k_a J^a (\vb \nu b, f^b) ,
\label{Ska}
\eeq
\beq
S_{\rm 2, \, LV}^{(k_\mu)} = \int d^4 x \, e \, \bar k_\mu J^\mu (\vb \nu b, f^\nu) .
\label{Skmu}
\eeq
In these terms, $f^b$ and $f^\nu$ are dynamical matter fields,
which are linked by the physical vierbein, obeying
\beq
f^\nu = \ivb \nu b f^b .
\label{fnuvb}
\eeq
The action $S_{\rm 1, \, LV}^{(k_a)}$ assumes $f^b$ are the basic matter field
components that are varied in order to obtain their equations of motion,
while $S_{\rm 2, \, LV}^{(k_\mu)}$ assumes $f^\nu$ are the basic field components.
The quantities $J^a$ and $J^\mu$ represent the parts of the Lagrangian terms that are contracted with $\bar k_a$
and $\bar k_\mu$, respectively.

The theories defined by these action terms are not the same.
This is because the background vierbein must be used to link
$\bar k_\mu$ and $\bar k_a$, 
while it is the dynamical vierbein that links $J^a$ and $J^\mu$.
As a result,
\bea
\bar k_\mu J^\mu &=& \bar e_\mu^{\pt{\mu} a} \bar k_a \, \ivb \mu b J^b 
\nonumber \\
&\ne& \bar k_a J^a  .
\eea
These action terms can also have different consequences concerning the
consistency conditions that must hold when the symmetry breaking is explicit.

For example, 
if an observer infinitesimal general coordinate transformation,
$x^\mu \rightarrow {x^\prime}^\mu = x^\mu - \xi^\mu$,
is performed it must leave the action unchanged in both cases,
since both theories are observer independent and have Lagrangians
that are observer scalars.
Mathematical identities that follow from this observer invariance
can then be obtained,
and these provide consistency conditions that must hold for each of these theories
\cite{rbas16}.

If these transformations are made in the first action followed by
Taylor expansions and relabeling,
the result is
\bea
0 = \de S_{\rm 1, \, LV}^{(\bar k_a)} = \int d^4 x \, \left(
\fr {\de (e \bar k_a J^a )} {\de \vb \mu b} {\cal L}_\xi \vb \mu b 
\right.
\nonumber \\
\left.
+ e \bar k_a \fr {\de J^a} {\de f^b} {\cal L}_\xi f^b
+ e J^a {\cal L}_\xi \bar k_a \right) ,
\label{diffSka}
\eea
where ${\cal L}_\xi$ are Lie derivatives.
Since $f^b$ is a dynamical field, the variations $\bar k_a \fr {\de J^a} {\de f^b}$ give the equations of motion for $f^b$,
which vanish on shell.  
Here,
\beq
{\cal L}_\xi \bar k_a = \xi^\nu \partial_\nu \bar k_a ,
\eeq
since $\bar k_a$ is a scalar on the spacetime manifold.
Using the definition of the energy-momentum tensor,
\beq
e T^{\mu\nu} = \fr {\de (e \bar k_a J^a )} {\de \vb \mu b} \uvb \nu b ,
\eeq
integrating by parts,
and putting the matter fields on shell, 
gives the result that
\beq
0 =  \int d^4 x \, e \left( - D_\mu T^{\mu}_{\pt{\mu}\nu}
+ J^a \partial_\nu \bar k_a \right) \xi^\nu .
\eeq
Since this must hold for all $\xi^\nu$ it follows that
\beq
D_\mu T^{\mu}_{\pt{\mu}\nu} = J^a \partial_\nu \bar k_a .
\label{nogo}
\eeq
Thus, in order for $D_\mu T^{\mu\nu} = 0$ to hold, 
which is required for consistency with the Bianchi identities,
it must be that 
\beq
J^a \partial_\nu \bar k_a = 0
\label{cond1}
\eeq
holds on shell over the spacetime manifold as a consistency condition.
Note that it was the severity of this
condition that led to the interpretation that 
explicit breaking is generally incompatible with Riemann
geometry in \cite{akgrav04}.

However,
with explicit breaking,
there are four extra degrees of freedom in the vierbein,
which would normally be gauged away in a theory with unbroken symmetry.
As long as these degrees of freedom do not decouple,
they can take values that satisfy \rf{cond1}.
This requires that the extra vierbein modes do not decouple in $J^a$,
which imposes a stringent condition on the theory.
If it turns out that the extra modes do decouple in $J^a$,
then the theory is incompatible with the Bianchi identity and
covariant energy-momentum conservation, 
and it must therefore be ruled out as a viable theory.

In contrast,
if similar procedures are followed starting with the second action,
$S_{\rm 2, \, LV}^{(\bar k_\mu)}$,
the resulting consistency conditions are not as stringent.
To see this,
infinitesimal general coordinate transformations 
followed by Taylor expansions and relabeling of coordinates
can again be performed.
The result in this case is
\bea
0 = \de S_{\rm 2, \, LV}^{(k_\mu)} = \int d^4 x \, \left(
\fr {\de (e \bar k_\si J^\si)} {\de \vb \mu a} {\cal L}_\xi \vb \mu a 
\right.
\nonumber \\
\left.
+ e \bar k_\mu \fr {\de J^\mu} {\de f^\nu} {\cal L}_\xi f^\nu
+ e J^\mu {\cal L}_\xi \bar k_\mu \right) ,
\label{diffSkmu}
\eea
Here,
the equations of motion for $f^\nu$ give
\beq
\bar k_\mu \fr {\de J^\mu} {\de f^\nu}=0 .
\eeq
However, in this case, $\bar k_\mu$ is a spacetime vector, 
and its Lie derivative is
\beq
{\cal L}_\xi \bar k_\mu = (D_\mu \xi^\nu) \bar k_\nu + \xi^\nu D_\nu \bar k_\mu .
\eeq
Using integration by parts, 
the result in this case is
\bea
0 =  \int d^4 x \, e \left( - D_\mu T^{\mu}_{\pt{\mu}\nu}
- (D_\mu J^\mu) \bar k_\nu \right.
\nonumber \\
\left. - (D_\mu \bar k_\nu ) J^\mu
+ J^\mu D_\nu \bar k_\mu \right) \xi^\nu .
\eea
Since this must hold for all $\xi^\nu$,
the result is the condition:
\beq
D_\mu T^{\mu}_{\pt{\mu}\nu} =
- (D_\mu J^\mu) \bar k_\nu 
- (D_\mu \bar k_\nu ) J^\mu
+ J^\mu D_\nu \bar k_\mu .
\label{nonogo}
\eeq
When $D_\mu T^{\mu\nu} = 0$,
consistency therefore requires that
\beq
- (D_\mu J^\mu) \bar k_\nu 
+ J^\mu (D_\nu \bar k_\mu - D_\mu \bar k_\nu) = 0
\label{identity}
\eeq
must hold.
Note that this consistency condition is different from the one in \rf {cond1},
and in general it is less restrictive.
This is because additional couplings to the metric appear
as a result of the covariant derivative $D_\mu J^\mu$ in \rf{identity},
and these extra degrees of freedom can take values that satisfy \rf{identity} 
even if the extra degrees of freedom decouple in $J^\mu$.
Thus, the action $S_{\rm 2, \, LV}^{(\bar k_\mu)}$ describes a theory that is
more generically compatible with the Bianchi identities than the one described by
$S_{\rm 1, \, LV}^{(\bar k_a)}$.

To summarize the results of this appendix,
a Lagrangian ${\cal L} = \bar k_a J^a$ is not the same as
one with ${\cal L} = \bar k_\mu J^\mu$,
and conclusions based on one of these forms do not apply for the other.
This is because $\bar k_a J^a \ne \bar k_\mu J^\mu$ when $\bar k_a$ and $\bar k_\mu$ are
components of a fixed background field.
Since the SME is defined in terms of
coefficients with spacetime indices,
such as $a_\mu$, $b_\mu$, $c_{\mu\nu}$, etc.,
it therefore has terms matching the form of ${\cal L} = \bar k_\mu J^\mu$,
where the SME coefficients replace $\bar k_\mu$.
In general, couplings of this type allow the extra degrees of freedom in the vierbein 
or metric to take values that satisfy the consistency conditions, 
and compatibility with the Bianchi identities and 
covariant energy-momentum conservation can therefore be maintained.
However, it is important to keep in mind that
if couplings to the local components of the SME coefficients are introduced,
the theory also needs to include couplings to the background vierbein $\barvb \mu a$
in order to maintain its overall consistency.  


\end{document}